\documentclass[11pt,a4paper]{article}
\usepackage[utf8]{inputenc}
\usepackage{amsmath}
\usepackage{amsfonts}
\usepackage{amssymb}
\usepackage{amsbsy}

\usepackage{graphicx}
\usepackage[left=3cm,right=3cm,top=2.5cm,bottom=2.5cm]{geometry}
\usepackage{algorithm, algorithmic}
\usepackage{mathrsfs}
\usepackage{hyperref}
\usepackage{tikz}
\usetikzlibrary{shapes,arrows}

\usepackage{multirow}
\usepackage{enumitem}
\usepackage{pdflscape}

\newcommand{\E}{\mathbb{E}}
\newcommand{\R}{\mathbb{R}}
\newcommand{\Z}{\mathbb{Z}}
\newcommand{\pr}{\mathbb{P}}
\newcommand{\vx}{\boldsymbol{x}}
\newcommand{\vy}{\boldsymbol{y}}
\newcommand{\vz}{\boldsymbol{z}}
\newcommand{\mX}{\boldsymbol{X}}
\newcommand{\mY}{\boldsymbol{Y}}
\newcommand{\mZ}{\boldsymbol{Z}}
\newcommand{\m}[1]{\boldsymbol{#1}}
\newcommand{\convr}[3]{\text{conv}_{\texttt{r}}^{\left(#1,#2,#3\right)}}
\newcommand{\convc}[2]{\text{conv}_{\texttt{c}}^{\left(#1,#2\right)}}
\newcommand{\attnr}[2]{\text{attn}_{\texttt{r}}^{\left(#1,#2\right)}}
\newcommand{\attns}[2]{\text{attn}_{\texttt{s}}^{\left(#1,#2\right)}}
\newcommand{\attnc}[3]{\text{attn}_{\texttt{c}}^{\left(#1,#2,#3\right)}}
\newcommand{\mlp}[2]{\text{mlp}^{\left(#1,#2\right)}}

\newcommand{\quantgan}[1]{\texttt{QuantGAN}}
\newcommand{\tagan}[1]{\texttt{TAGAN}}
\newcommand{\ttgan}[1]{\texttt{TTGAN}}
\newcommand{\gans}[1]{{GANs}}
\newcommand{\gan}[1]{{GAN}}
\newcommand{\wgangp}[1]{{WGAN-GP}}
\newcommand{\garch}[1]{{GARCH}}

\title{\textbf{Simulating financial time series using attention}}
\date{ }
\author{Weilong Fu\footnote{Department of IEOR, Columbia University, \url{wf2232@columbia.edu}}, Ali Hirsa\footnote{Department of IEOR, Columbia University, \url{ah2347@columbia.edu}}, J{\"o}rg Osterrieder\footnote{School of Engineering, Zurich University of Applied Sciences, \url{oste@zhaw.ch}}}

\begin{document}
\maketitle

\abstract{Financial time series simulation is a central topic since it extends the limited real data for training and evaluation of trading strategies. It is also challenging because of the complex statistical properties of the real financial data. We introduce two generative adversarial networks (\gans{}), which utilize the convolutional networks with attention and the transformers, for financial time series simulation. The \gans{} learn the statistical properties in a data-driven manner and the attention mechanism helps to replicate the long-range dependencies. The proposed \gans{} are tested on the S\&P 500 index and option data, examined by scores based on the stylized facts and are compared with the pure convolutional \gan{}, i.e. \quantgan{}. The attention-based \gans{} not only reproduce the stylized facts, but also smooth the autocorrelation of returns.}

\providecommand{\keywords}[1]{{\textit{Keywords:}} #1}
\keywords{deep learning, generative adversarial networks, attention, time series, stylized facts}

\section{Introduction}
Training and evaluation of trading strategies need lots of data. Due to the limited amount of real data, there is a growing need to be able to simulate realistic financial data which satisfies the stylized facts. There has already been a vast literature of financial time series models. The generalized autoregressive conditional heteroskedasticity (\garch{}) \cite{bollerslev_generalized_1986} model and its variants are applied to the stock prices and indices. The Black-Merton-Scholes model \cite{black_scholes_1973}, the Heston model \cite{heston1993closed}, the variance gamma model \cite{madan1990variance}, etc. are applied to the option surfaces. The parametric models are popular for their simplicity, mathematical explicitness and robustness. However, it is difficult for a parametric model to fit all the major stylized facts. 

Recently, more data-driven approaches based on generative adversarial networks (\gans{}) \cite{goodfellow2014generative} are proposed to deal with the problem. The \gan{} includes a generator, which is used to generate samples, and a discriminator, which is responsible for judging whether the generated samples are similar enough to the real data. The applications of \gans{} to financial time series range from the underlying asset price prediction \cite{zhou_stock_2018,zhang_stock_2019,koshiyama_generative_2020} and simulation \cite{takahashi_modeling_2019,de2019enriching,kondratyev_market_2019,fu2019time,wiese_quant_2020} to the option surface simulation \cite{wiese2019deep}. Some more \gans{} of time series are proposed in \cite{vandenoord16_ssw,mogren2016c,esteban2017real,che2018hierarchical,yoon_time-series_2019,koochali_gan_2019} and some more generative models of time series are in \cite{hofert_multivariate_2020,ni_conditional_2020,wiese2021multi}. However, the network structures of the \gans{} in financial time series simulation are mostly limited in convolutional networks \cite{lecun_backpropagation_1989} and recurrent networks \cite{hochreiter1997long,chung2014empirical}. 

There have been various different \gans{} which employ the attention mechanism \cite{bahdanau2014neural,luong2015effective} to improve their performance, e.g., the convolutional networks with attention \cite{zhang_sagan_2019,brock2018large,daras_2020_CVPR}, and the transformer networks \cite{jiang_gan_2021,hudson_gan_2021}. But the attention mechanism has not been tested on financial time series. It is shown in \cite{wiese_quant_2020} that long-range dependency is a major challenge in financial time series simulation. The attention mechanism \cite{bahdanau2014neural,luong2015effective} is perfectly suitable for modeling the stylized facts of long-range dependency due to the large receptive field size of the attention layer. Thus we are motivated to use the attention-based \gans{} to simulate the financial time series. It is important to note the difference between time series and fixed-dimensional variables, since a time series can have an arbitrary length. Thus we need to modify the attention-based \gans{} to make it agnostic to the length of the time series by changing all the layers in the generator to the causal layers. Our findings based on numerical results show that the attention-based \gans{} perform as well as the temporal convolutional network (WaveNet \cite{vandenoord16_ssw}) in replication of the major stylized facts, including heavy tails, autocorrelation and cross-correlation, and are better at simulating smooth autocorrelation of returns and satisfying the no-arbitrage condition of option surfaces. It is well known in the literature that both \gans{} and transformers \cite{vaswani_attention_17}, which contain multiple attention layers, are hard to train, and it is a challenging problem to combine them together. Millions or billions of samples are usually used to train the transformer \gans{} in text generation and image simulation. Authors in \cite{jiang_gan_2021} pointed it out that transformers are data-hungry, and thus they made used of data augmentation techniques to improve the transformer \gans{}. In this paper, we propose a new transformer \gan{} using sparse attention \cite{child2019generating,daras_2020_CVPR} and train it using a small amount of financial time series data (around 3000 samples).  

The rest of this paper is organized as follows. In Section \ref{sec:model}, we define the generative model and emphasize the difference between the generative models of time series and fixed-dimensional distributions. We also introduce the \gans{}, which employ a discriminator to train the generator. In Section \ref{sec:layer}, we introduce and compare the regular and causal convolutional layers and attention layers, which are building blocks of the proposed \gans{}. In Section \ref{sec:network}, we propose to employ attention as the tool to model the long-range dependencies, and give the detailed structure of the proposed temporal attention \gan{} (\tagan{}) and temporal transformer \gan{} (\ttgan{}). In Sections \ref{sec:index_simulation} and \ref{sec:option_simulation}, we show the numerical results of the proposed \gans{} for the S\&P 500 index simulation and its index option surface simulation respectively. Most graphical results are collected in Appendix \ref{app:results}. Section \ref{sec:conclusion} summarizes the paper.

\section{Generative model of financial time series}\label{sec:model}
\subsection{Problem formulation}
Suppose we have a financial time series $\{\vx_t\in \R^d\}_{t\in\Z}$, e.g., the historical of prices or volatilities, and would like to generate a time series $\{\vy_t\in \R^d\}_{t\in\Z}$ that has the same statistical properties given a series of i.i.d. random noise $\{\vz_t\in\R^{d_n}\}_{t\in\Z}$ via deep learning. Let $\vz_{i:j}$ denote the sequence $\{\vz_i,\vz_{i+1},\dots,\vz_{j}\}$ and the same notation is used for all time series hereafter.

Here we follow the definition of the \textit{neural process} in \cite{wiese_quant_2020}. We would like to develop a generator $G(\cdot;\theta_G)$, a neural network with the parameter $\theta_G\in\Theta_G$, which takes random noise from $\{\vz_t\}_{t\in\Z}$ as its input and outputs the time series $\{\vy_t\}_{t\in\Z}$, i.e., $$\vy_t = G(\vz_{t-f+1:t};\theta_G),\forall t\in\Z$$ where $f>0$ is called the receptive field size (RFS) and means the length of noise variables of which $\vy_t$ is composed. By this definition, $\vy_t$ and $\vy_{t+\tau}$ would be independent if $\tau\geq f$. Also, since $\vy_t$ is computed from $\{\vz_\tau\}_{\tau\leq t}$, we know $\{\vy_t\}_{t\in\Z}$ is adapted to $\{\vz_t\}_{t\in\Z}$.

The generator of time series has to satisfy the following conditions that make it different from a generator of a fixed-dimensional distribution. 
\begin{itemize}
	\item[(a)] The generator should be able to take in random noise of length $l+f-1$ to output the aimed time series of an arbitrary length $l$, i.e.,
	$$\vy_{t-l+1:t}= G(\vz_{t-l-f+2:t};\theta_G),\forall t\in\Z.$$
	\item[(b)] The generated time series can be prolonged in a consistent manner. For arbitrary $t_1,t_2,t_3,t_4\in\Z$ such that $[t_1,t_2]\cap[t_3,t_4]\neq \emptyset$, $$\vy_{t_1:t_2}= G(\vz_{t_1-f+1:t_2};\theta_G)$$ and $$\tilde\vy_{t_3:t_4}= G(\vz_{t_3-f+1:t_4};\theta_G),$$ the overlapping part of the generated time series must be equal, i.e., $$\vy_{\max\{t_1,t_3\}:\min\{t_2,t_4\}}=\tilde\vy_{\max\{t_1,t_3\}:\min\{t_2,t_4\}}.$$ This means the generated time series $\{\vy_t\}_{t\in\Z}$ is uniquely determined by the random noise series $\{\vz_t\}_{t\in\Z}$.
\end{itemize}
Given the two conditions are satisfied, we can always let the generator $G(\cdot;\theta_G)$ compute sequences of length $l$ and then combine the sequences to make up a longer sequence $\vy_{1:T}$, where $T>l$. To be more specific, we first calculate the pieces 
$$\vy_{(i-1)l+1:il}= G(\vz_{(i-1)l-f+2:il};\theta_G),\forall 1\leq i\leq \lceil T/l\rceil,$$
and then $\vy_{1:T}$ is a subsequence of $\vy_{1:l\lceil T/l\rceil}$, where $\lceil\cdot\rceil$ is the ceiling function.

\subsection{Training through generative adversarial network}
We give a quick introduction of generative adversarial networks (\gans{}) as well as the loss functions for training \gans{}. The \gan{} introduces a discriminator $$D(\cdot;\theta_D):\R^{l\times d}\rightarrow \R,$$ where $\theta_D\in\Theta_D$, to evaluate the similarity between the real historical data $\{\boldsymbol{x}_t\}_{t\in\Z}$ and the simulated data $\{\boldsymbol{y}_t\}_{t\in\Z}$. A higher output value from $D(\cdot;\theta_D)$ means the discriminator holds a stronger belief that the input sample comes from the real data. 

Suppose we have a sequence of real data $\vx_{1:T}$. Let $\pr_{\mX}$ be the uniform distribution over the window data of length $l$, $\{\vx_{i:i+l-1},\forall 1\leq i\leq T-l+1\}$. Also, let $\pr_{\mZ}$ be distribution of $\vz_{1:l+f-1}$. Then we draw $\mX\sim\pr_{\mX}$ to be a piece of real data of length $l$ and $\mZ\sim\pr_{\mZ}$ the random noise of length $l+f-1$ and get the simulated sequence $\mY=G(\mZ;\theta_G)\in \R^{l\times {d}}$. The \gan{} trains the generator and the discriminator by minimizing the following loss functions
$$\min_{\theta_G} \E_{\mZ} \mathcal{L}_G(D(G(\mZ;\theta_G);\theta_D))$$
and
$$\min_{\theta_D} \E_{\mX, \mZ,\tilde{\mX}} \mathcal{L}_D(D(\mX;\theta_D),D(G(\mZ;\theta_G);\theta_D),\nabla_{\tilde{\mX}}D(\tilde{\mX};\theta_D)),$$
where $\mathcal{L}_G(\cdot)$ and $\mathcal{L}_D(\cdot,\cdot,\cdot)$ are the loss functions of the discriminator and the generator. The third argument in $\mathcal{L}_D(\cdot,\cdot,\cdot)$ is not included in the original \gan{} but related with gradient penalty, where $\tilde{\mX}=(1-U)\mX+U\mY$ is a linear interpolation between $\mX$ and $\mY$ requiring $U$ follows the uniform distribution over $(0,1)$.

The loss functions of the original \gan{} \cite{goodfellow2014generative} are  
\begin{align}
\begin{split}
	\mathcal{L}_G(d_f)&= -\ln(\sigma(d_f))\\
	\mathcal{L}_D(d_r,d_f,\boldsymbol{g})&= -\ln(\sigma(d_r))-\ln(1-\sigma(d_f))
\end{split}\label{eq:loss_gan}
\end{align} 
where $\sigma(d)=1/(1+e^{-d})$ is the sigmoid function and $\sigma(D(\mX;\theta_D))$ means the probability that the discriminator considers $\mX$ belongs to the real data. A quick derivation of the losses is included in Appendix \ref{app:gan_loss}. 

Besides the original losses, the loss functions of the Wasserstein \gan{} \cite{arjovsky_wasserstein_2017} with gradient penalty (\wgangp{}) \cite{gulrajani2017improved} are also widely used, where the gradient norm penalty is used to achieve Lipschitz continuity: 
\begin{align}
\begin{split}
    \mathcal{L}_G(d_f) &= - d_f\\
    \mathcal{L}_D(d_r,d_f,\boldsymbol{g}) &= -d_r + d_f + \lambda (\Vert \boldsymbol{g} \Vert-1)^2
\end{split}\label{eq:loss_wgan_gp}
\end{align}
where $\lambda$ is a constant and $\lambda=10$ by default. $\Vert \cdot \Vert$ is the Frobenius norm. In Appendix \ref{app:gan_loss}, we introduce how the losses are derived.

\section{Network layers}\label{sec:layer}
In this section, we going to list all the layers that will be used in the proposed network structure. Some of the layers are already introduced in literature, but we still give a short introduction for each to make the paper self-contained. The layers are classified into regular layers and causal layers. In the causal layers, each output node only depends on the input nodes with equal or smaller time indices. Suppose the input of the causal layer is $\m{I}\in\R^{n_l\times n_i}$ with rows $\{\m{I}_{t,\cdot}\}_{t=1}^{n_l}$ and the output is $\m{O}\in\R^{(n_l-f+1)\times n_o}$ with rows $\{\m{O}_{t,\cdot}\}_{t=f}^{n_l}$, then each row of the output $\m{O}_{t,\cdot}$ only depends on $\{\m{I}_{\tau,\cdot}\}_{\tau=t-f+1}^{t}$. However, the regular layers are not subject to this restriction. Since the output of the generator needs to be adapted to the input noise, the causal layers are used in the generator. While the regular layers admit more flexibility and are used in the discriminator. 
 
\subsection{Regular convolutional layer}
\begin{figure}[h]
\centering
	\begin{tikzpicture}[shorten >=1pt,->,draw=black!50, node distance=2.5 cm]
    \tikzstyle{every pin edge}=[-,shorten <=1pt]
    \tikzstyle{neuron}=[circle,fill=black!25,minimum size=24pt,inner sep=0pt]
    \tikzstyle{input neuron}=[neuron];
    \tikzstyle{output neuron}=[neuron];
    \tikzstyle{hidden neuron}=[neuron];
    \tikzstyle{annot} = [text width=6em, text centered]

    \foreach \name / \y in {1,...,6}
        \node[input neuron, pin=above:Input] (I-\name) at (2*\y,0) {$\m{I}_{\name,\cdot}$};
    \foreach \name / \y in {1,...,6}
        \node[hidden neuron, pin=below:Output] (O-\name) at (2*\y,-2cm) {$\m{O}_{\name,\cdot}$};
    \foreach \x in {1,...,6}
            \path (I-\x) edge (O-\x);
	\foreach \x[evaluate=\x as \evalx using int(\x+1)] in {1,...,5}{
	    \path (I-\x) edge (O-\evalx);}
	\foreach \x[evaluate=\x as \evalx using int(\x-1)] in {2,...,6}{
	    \path (I-\x) edge (O-\evalx);}
\end{tikzpicture}
\caption{Illustration of the regular convolutional layer (length $n_l=6$, kernel size $n_k=3$ and stride $s=1$).}
\label{fig:regu_conv}
\end{figure}
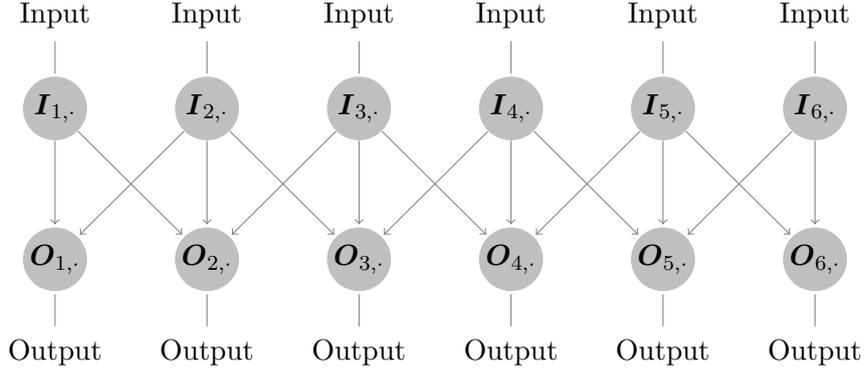

In \cite{lecun_backpropagation_1989}, the authors proposed the convolutional layer, which is good at extracting local information. The two-dimensional case is widely used in computer vision and the one-dimensional case is used in sequence models. Although the convolutional layer is widely used and well-known, we reiterate the definition to show the difference between the different layers. Suppose the input is $\m{I}\in\R^{n_l\times n_i}$ and it passes through a one-dimensional regular convolutional layer with kernel size $n_k$, output channel $n_o$ and stride $s$. The kernel size $n_k$ is an odd number by default. The parameters are the weight $\m{W}\in\R^{n_k\times n_i\times n_o}$ and the intercept $\m{b}\in\R^{n_o}$. The output of the regular convolutional layer is $\m{O}\in\R^{\lfloor n_l/s\rfloor\times n_o}$ given by 
\begin{align*}
	{O}_{i_l,i_o}=\sum_{{i}=1}^{n_i}\sum_{i_k=1}^{n_k}W_{i_k,{i},i_o}I_{s(i_l-1)+1-(n_k+1)/2+i_k,{i}} + b_{i_o},\forall 1\leq i_l\leq \lfloor n_l/s\rfloor,1\leq i_o\leq n_o,
\end{align*}
where $\lfloor\cdot\rfloor$ is the floor function. The `same' padding rule is applied to the input, i.e., $I_{i_l,{i}}=I_{1,{i}},\forall i_l<1$ and $I_{i_l,{i}}=I_{n_l,{i}},\forall i_l>n_l$. The regular convolutional layer is illustrated in Figure \ref{fig:regu_conv}. It is denoted as $\convr{n_k}{n_o}{s}(\cdot)$, where $n_k,n_o$ and $s$ are the kernel size, output channel and stride respectively.

\subsection{Causal convolutional layer}
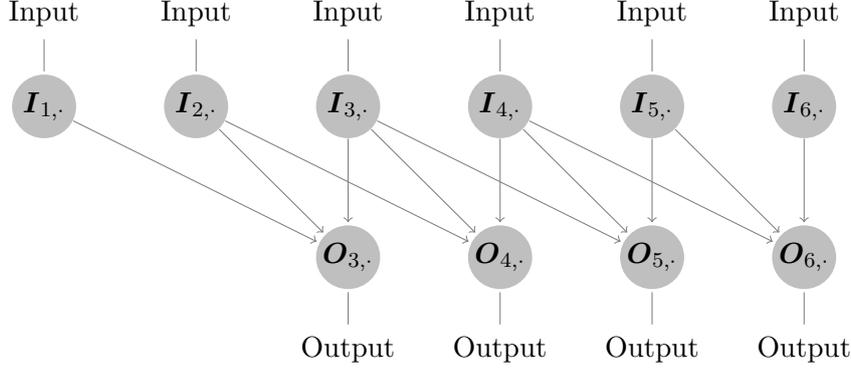
\begin{figure}[h]
\centering
	\begin{tikzpicture}[shorten >=1pt,->,draw=black!50, node distance=2.5 cm]
    \tikzstyle{every pin edge}=[-,shorten <=1pt]
    \tikzstyle{neuron}=[circle,fill=black!25,minimum size=24pt,inner sep=0pt]
    \tikzstyle{input neuron}=[neuron];
    \tikzstyle{output neuron}=[neuron];
    \tikzstyle{hidden neuron}=[neuron];
    \tikzstyle{annot} = [text width=6em, text centered]

    \foreach \name / \y in {1,...,6}
        \node[input neuron, pin=above:Input] (I-\name) at (2*\y,0) {$\m{I}_{\name,\cdot}$};
    \foreach \name / \y in {3,...,6}
        \node[hidden neuron, pin=below:Output] (O-\name) at (2*\y,-2cm) {$\m{O}_{\name,\cdot}$};
    \foreach \x in {3,...,6}
            \path (I-\x) edge (O-\x);
	\foreach \x[evaluate=\x as \evalx using int(\x+1)] in {2,...,5}{
	    \path (I-\x) edge (O-\evalx);}
	\foreach \x[evaluate=\x as \evalx using int(\x+2)] in {1,...,4}{
	    \path (I-\x) edge (O-\evalx);}
\end{tikzpicture}
\caption{Illustration of the causal convolutional layer (length $n_l=6$ and kernel size $n_k=3$).}
\label{fig:causal_conv}
\end{figure}

In \cite{vandenoord16_ssw}, the authors proposed the causal convolutional layer to model the audio data. 
%The dilated causal convolutional layer can build a network with a large RFS, which was strictly defined in \cite{wiese_quant_2020}. However, since we use the attention layer to increase the RFS, we only need the causal convolutional layer without dilation. 
Suppose the input is $\m{I}\in\R^{n_l\times n_i}$ and it passes through a causal convolutional layer with kernel size $n_k$ and output channel $n_o$. The parameters are the weight $\m{W}\in\R^{n_k\times n_i\times n_o}$ and the intercept $\m{b}\in\R^{n_o}$. The output of the causal convolutional layer is $\m{O}\in\R^{(n_l-n_k+1)\times n_o}$. In the causal layer, the time index of the output is taken from $\{n_k,n_k+1,\dots,n_l\}$. The output is given by 
\begin{align*}
	{O}_{t,i_o}=\sum_{{i}=1}^{n_i}\sum_{i_k=1}^{n_k}W_{i_k,{i},i_o}I_{t-n_k+i_k,{i}} + b_{i_o},\forall n_k\leq t\leq n_l,1\leq i_o\leq n_o.
\end{align*}
The RFS of the causal convolutional layer is equal to $n_k$. The causal convolutional layer is illustrated in Figure \ref{fig:causal_conv}. It is denoted as $\convc{n_k}{n_o}(\cdot)$, where $n_k$ and $n_o$ are the kernel size and output channel.

\subsection{Regular attention layer}
\begin{figure}[h]
\centering
	\begin{tikzpicture}[shorten >=1pt,->,draw=black!50, node distance=2.5 cm]
    \tikzstyle{every pin edge}=[-,shorten <=1pt]
    \tikzstyle{neuron}=[circle,fill=black!25,minimum size=24pt,inner sep=0pt]
    \tikzstyle{input neuron}=[neuron];
    \tikzstyle{output neuron}=[neuron];
    \tikzstyle{hidden neuron}=[neuron];
    \tikzstyle{annot} = [text width=6em, text centered]

    \foreach \name / \y in {1,...,6}
        \node[input neuron, pin=above:Input] (I-\name) at (2*\y,0) {$\m{I}_{\name,\cdot}$};
    \foreach \name / \y in {1,...,6}
        \node[hidden neuron, pin=below:Output] (O-\name) at (2*\y,-2cm) {$\m{O}_{\name,\cdot}$};
    \foreach \x in {1,...,6}
    	\foreach \y in {1,...,6}
            \path (I-\x) edge (O-\y);
\end{tikzpicture}
\caption{Dependence relationship in the regular attention layer (length $n_l=6$).}
\label{fig:regu_attn}
\end{figure}
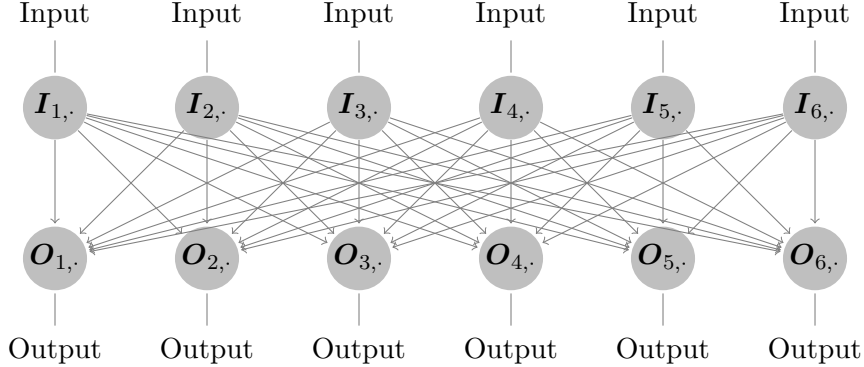

The attention layer was introduced by \cite{bahdanau2014neural,luong2015effective}. It is designed to receive and process global information to improve the performance of convolutional or recurrent networks. The attention layer was first used in text translation, where input words and output words may follow different orders because of different grammar. So, the attention layer needs to search from the entire inputs to decide which input word is corresponding to a specific output word. The best match in the input becomes the `attention' of the layer. Authors in \cite{vaswani_attention_17} proposed the transformer network, which consists of only attention layers and multi-layer perceptrons, and proved attention layers are capable of modeling sequences without help from convolutional or recurrent layers. Suppose the input of the regular multi-head attention layer is $\m{I}\in\R^{n_l\times n_i}$, the hidden size is $n_a$ and the number of heads is $n_h$, which satisfy mod$(n_a,n_h)=0$. The parameters are the weights $\m{W}^{Q},\m{W}^{K},\m{W}^{V}\in \R^{n_i\times n_a}$, $\m{W}^{O}\in \R^{n_a\times n_i}$ and intercepts $\m{b}^{Q},\m{b}^{K},\m{b}^{V}\in \R^{n_a}$, $\m{b}^{O}\in \R^{n_i}$. Let $\m{1}$ be the vector of length $n_l$ with all elements of 1. The formulae in the regular attention layer are
\begin{align}
\begin{split}
	\m{Q} &= \m{I}\m{W}^{Q}+\m{1}{\m{b}^{Q}}^{\top}\\
	\m{K} &= \m{I}\m{W}^{K}+\m{1}{\m{b}^{K}}^{\top}\\
	\m{V} &= \m{I}\m{W}^{V}+\m{1}{\m{b}^{V}}^{\top}\\
	\m{A}_{(i_h)} &= \text{softmax}\left(\m{Q}_{(i_h)}\m{K}_{(i_h)}^{\top}\right)\m{V}_{(i_h)},\forall 1\leq i_h\leq n_h\\
	\m{O} &= \m{A}\m{W}^{O}+\m{1}{\m{b}^{O}}^{\top}
\end{split}\label{eq:regu_attn}
\end{align}
where $\m{A}_{(i_h)}$ means the submatrix from column $(i_h-1)n_h+1$ to column $i_h n_h$ and $\m{O}\in\R^{n_l\times n_i}$ is the output of the attention layer. The softmax function is evaluated along each row of the input matrix. While the fourth equation in Equation \eqref{eq:regu_attn} is often replaced with 
\begin{align*}
	\m{A}_{(i_h)} &= \text{softmax}\left(\m{Q}_{(i_h)}\m{K}_{(i_h)}^{\top}/\sqrt{n_a/n_h}\right)\m{V}_{(i_h)}
\end{align*}
when the size of a single attention head $n_a/n_h$ is large, we stick with the equation in Equation \eqref{eq:regu_attn} since it shows better empirical results when $n_a/n_h$ is small. The dependence relationship of the output on the input in the regular attention layer is illustrated in Figure \ref{fig:regu_attn}. The regular attention layer is denoted as $\attnr{n_a}{n_h}(\cdot)$, where $n_a$ and $n_h$ are the hidden size and number of heads.

\subsection{Sparse attention layer}
The sparse attention layer is introduced by \cite{child2019generating} to accelerate computation and improve the focus of the attention layer. The sparse attention introduces the sparse masks before the softmax function, i.e. replacing the fourth equation in Equation \eqref{eq:regu_attn} with 
\begin{align}
	\m{A}_{(i_h)} &= \text{softmax}\left(\m{Q}_{(i_h)}\m{K}_{(i_h)}^{\top}+\m{M}^{\left(i_h\right)} \right)\m{V}_{(i_h)}
\end{align}
where $\m{M}^{\left(i_h\right)}\in \R^{n_i\times n_i},\forall 1\leq i_h\leq n_h$ are the sparse mask matrices. In the mask $\m{M}^{\left(i_h\right)}$, the elements in the masked positions are assigned a large negative value $-L$, while the elements outside the mask are assigned 0. Since they are sparse masks, the 0 elements are sparse and the large negative value elements are dense. The masked positions are mapped to 0 by the softmax function since $e^{-L}\approx 0$, where we usually let $L=10^{3}$. In this way, the masked positions in $\m{Q}_{(i_h)}\m{K}_{(i_h)}^{\top}$ are not involved in the result of the softmax function and the attention is limited within the sparse mask.

\begin{figure}[h]
    \centering
 	\includegraphics[width = 0.75\textwidth]{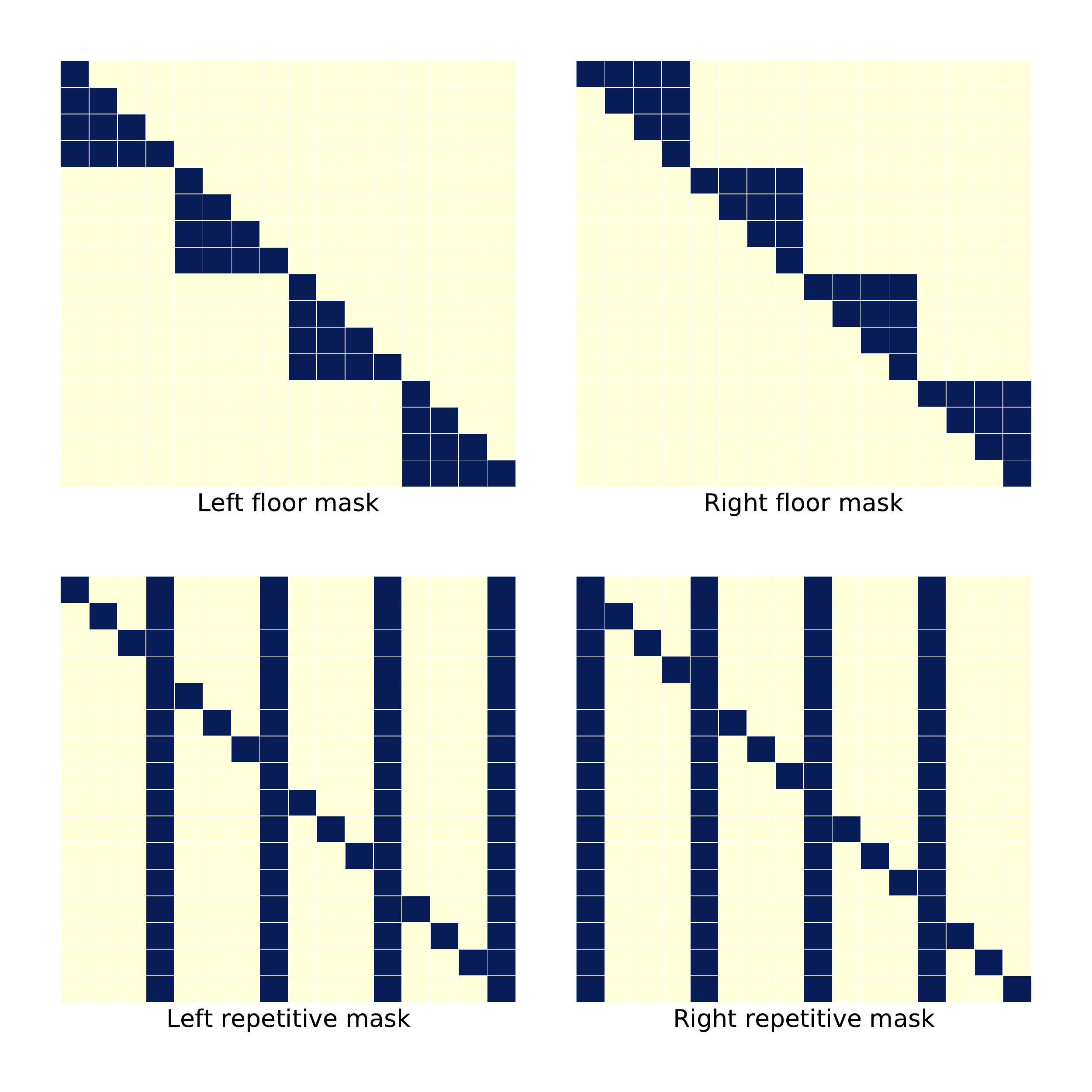}
 	\caption{Example of the sparse masks when $n_l=16$. Light elements are masked while dark elements are not.}
    \label{fig:sparse_mask}
\end{figure}

In this paper we use the sparse masks proposed in \cite{daras_2020_CVPR}. They are generated in the following way. Let $s=\lfloor\sqrt{n_l}\rfloor$ be the stride, where $\lfloor\cdot\rfloor$ is the floor function. We then create the index sets for the masks:
\begin{itemize}
	\item Left floor mask: $S_1 = \{(i,j):\lfloor(i-1)/s\rfloor=\lfloor(j-1)/s\rfloor\,\text{and}\,i\geq j\}$
	\item Right floor mask: $S_2 = \{(i,j):\lfloor(i-1)/s\rfloor=\lfloor(j-1)/s\rfloor\,\text{and}\,i\leq j\}$
	\item Left repetitive mask: $S_3 = \{(i,j):\text{mod}(j,s)=0\,\text{or}\,i=j\}$
	\item Right repetitive mask: $S_4 = \{(i,j):\text{mod}(j,s)=1\,\text{or}\,i=j\}$
\end{itemize}
The corresponding masks are defined as 
\begin{align*}
	{M}^{(i_h)}_{i,j}=\begin{cases}
		0,&\text{if}\,(i,j)\in S_{i_s},\\
		-L,&\text{if}\,(i,j)\notin S_{i_s},  
	\end{cases}\,\text{when}\,i_h\equiv i_s (\text{mod}\,4).
\end{align*}
As a result, the number of heads $n_h$ needs to be a multiple of 4 when we use the sparse attention. The sparse attention limits the attention of each node to a specific region such that the network would converge faster. In Figure \ref{fig:sparse_mask}, we show an example of the sparse masks. As shown in Figure \ref{fig:sparse_mask}, the left and right floor masks focus on local features and the left and right repetitive masks focus on periodic features. The sparse attention layer is denoted as $\attns{n_a}{n_h}(\cdot)$, where $n_a$ and $n_h$ are the hidden size and number of heads.

\subsection{Causal attention layer}
\begin{figure}[h]
\centering
	\begin{tikzpicture}[shorten >=1pt,->,draw=black!50, node distance=2.5 cm]
    \tikzstyle{every pin edge}=[-,shorten <=1pt]
    \tikzstyle{neuron}=[circle,fill=black!25,minimum size=24pt,inner sep=0pt]
    \tikzstyle{input neuron}=[neuron];
    \tikzstyle{output neuron}=[neuron];
    \tikzstyle{hidden neuron}=[neuron];
    \tikzstyle{annot} = [text width=6em, text centered]

    \foreach \name / \y in {1,...,6}
        \node[input neuron, pin=above:Input] (I-\name) at (2*\y,0) {$\m{I}_{\name,\cdot}$};
    \foreach \name / \y in {4,...,6}
        \node[hidden neuron, pin=below:Output] (O-\name) at (2*\y,-2cm) {$\m{O}_{\name,\cdot}$};
    \foreach \x in {1,...,4}
    	\path (I-\x) edge (O-4);
    \foreach \x in {2,...,5}
    	\path (I-\x) edge (O-5);
    \foreach \x in {3,...,6}
    	\path (I-\x) edge (O-6);
\end{tikzpicture}
\caption{Dependence relationship in the causal attention layer (length $n_l=6$ and RFS $n_f=4$).}
\label{fig:causal_attn}
\end{figure}
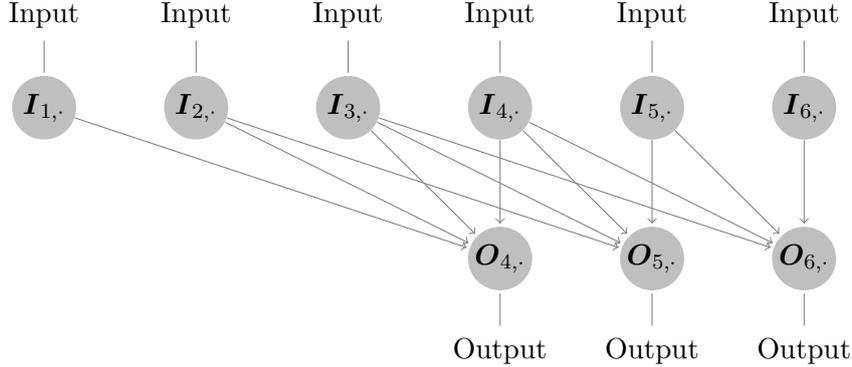

To make use of the attention layers in the generator of time series, we need to make sure the output only depends on the input elements in the past. This is also achieved by using masks. We suppose the same input $\m{I}\in\R^{n_l\times n_i}$ as before, and the parameters are defined the same as in the regular attention. Let the RFS of the layer be $n_f$. The definition of the attention layer in Equation \eqref{eq:regu_attn} needs to be replaced with 
\begin{align*}
	\m{Q} &= \m{I}\m{W}^{Q}+\m{1}{\m{b}^{Q}}^{\top}\\
	\m{K} &= \m{I}\m{W}^{K}+\m{1}{\m{b}^{K}}^{\top}\\
	\m{V} &= \m{I}\m{W}^{V}+\m{1}{\m{b}^{V}}^{\top}\\
	\m{A}_{(i_h)} &= \text{softmax}\left(\m{Q}_{(i_h)}\m{K}_{(i_h)}^{\top}+\m{M}\right)\m{V}_{(i_h)},\forall 1\leq i_h\leq n_h\\
	\m{O} &= \left(\m{A}\m{W}^{O}+\m{1}{\m{b}^{O}}^{\top}\right)_{n_f:n_l,\cdot}
\end{align*}
where $(\cdot)_{n_f:n_l,\cdot}$ means the submatrix from row $n_f$ to row $n_l$, and $\m{M}$ is a mask matrix with the elements $$M_{i,j}=\begin{cases}
	0,&\text{if}\,\,\,0\leq i-j\leq n_f-1,\\
	-L,&\text{else}.
\end{cases}$$
In this way, each output only depends on the current input and $n_f-1$ past inputs. The benefit of the causal attention layer is that it can increase the RFS $n_f$ arbitrarily without introducing additional parameters. The size of the parameters does not depend on the input length $n_l$ or the RFS $n_f$. The dependence relationship of the output on the input in the causal attention layer is illustrated in Figure \ref{fig:causal_attn}. The causal attention layer is denoted as $\attnc{n_a}{n_h}{n_f}(\cdot)$, where $n_a,n_h$ and $n_f$ are the hidden size, number of heads and RFS respectively.

\subsection{Multi-layer perceptron block}
The multi-layer perceptron (MLP) is a key component of the transformer architecture. Suppose the input is $\m{I}\in\R^{n_l\times n_i}$, the hidden size is $n_m$ and the activation function is $h(\cdot)$. The parameters are the weights $\m{W}^{(1)}\in \R^{n_i\times n_m}$, $\m{W}^{(2)}\in \R^{n_m\times n_i}$ and intercepts $\m{b}^{(1)}\in\R^{n_m}$, $\m{b}^{(2)}\in \R^{n_i}$. Let $\m{1}$ be the vector of length $n_l$ with all elements of 1. The formulae in the MLP block are
\begin{align*}
	\m{H} &= \m{I}\m{W}^{(1)}+\m{1}{\m{b}^{(1)}}^{\top}\\
	\m{O} &= h(\m{H})\m{W}^{(2)}+\m{1}{\m{b}^{(2)}}^{\top}
\end{align*}
where the activation function $h(\cdot)$ is applied element-wise and $\m{O}\in\R^{n_l\times n_i}$ is the output. All the activation functions will be applied element-wise in the paper. The MLP block is denoted as $\mlp{n_m}{h}(\cdot)$, where $n_m$ and $h(\cdot)$ are the hidden size and the activation function respectively.

\section{Network structures}\label{sec:network}
\subsection{Need for a large receptive field size}
\begin{figure}[h]
    \centering
 	\includegraphics[width = 0.6\textwidth]{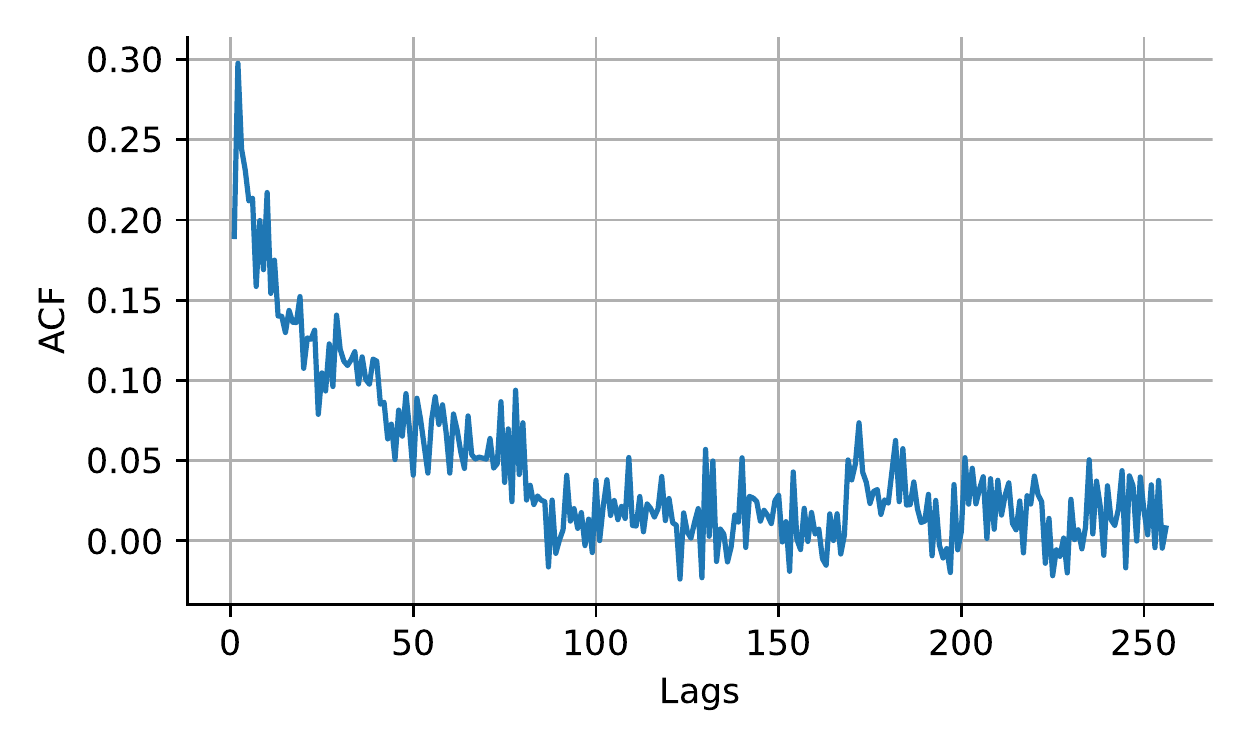}
 	\caption{ACF of absolute values of the S\&P 500 index returns.}
    \label{fig:abs_acf}
\end{figure}

The difficulty of financial time series simulation is to model the long-range dependencies. Figure \ref{fig:abs_acf} shows the autocorrelation of absolute values of the returns of the S\&P 500 index from May 2010 to November 2018. The autocorrelation is positive, inferring the asset returns admit phases of high activity and low activity in terms of price changes. This stylized fact is called volatility clustering. The positive correlation decays to almost 0 when the lag is greater than 100. This means we need a generator of RFS larger than 100 to model the positive correlation in this data. Authors in \cite{vandenoord16_ssw,wiese_quant_2020} make use of the temporal convolutional network (TCN) to increase the RFS. While in this paper we use the attention layer to build a generator of a large RFS.

\subsection{Temporal attention \gan{}}
The temporal attention \gan{} (\tagan{}) is composed of a generator and a discriminator which are both convolutional networks with one self-attention layer. In the generator, we always use the causal layers while in the discriminator we only use the regular layers. This proposed model is the modification to the self-attention \gan{} \cite{zhang_sagan_2019} which has shown good performance in image generation. The hyper-parameters of \tagan{} are listed in Table \ref{tab:tagan_paras}.
\begin{table}[h]
\begin{small}
\centering
	\begin{tabular}{rl}\hline
	hyper-parameter & meaning \\ \hline
	$l$ & data length\\
	$f$ & receptive field size \\
	$d_n$ & noise channel \\
	$d$ & data channel \\
	$d_h$ & hidden channel in the generator\\
	$d_s$ & start hidden channel in the discriminator\\
	$d_m$ & max hidden channel in the discriminator\\
	$n_k$ & kernel size in convolutions\\ 
	$L_{1}^{G}$/$L_{2}^{G}$ & number of convolutional blocks before/after attention in the generator\\
	$L_{1}^{D}$/$L_{2}^{D}$ & number of convolutional blocks before/after attention in the discriminator\\
	$n_h$ & number of heads in attention\\
	$n_{a,G}$/$n_{a,D}$ & attention hidden size in the generator/discriminator\\
	$h_G(\cdot)$/$h_D(\cdot)$ & activation function in the generator/discriminator\\
	\hline
	\end{tabular}
\end{small}
	\caption{Hyper-parameters in \tagan{}.}
	\label{tab:tagan_paras}
\end{table}

\begin{figure}[h]
\centering
\begin{tikzpicture}[x=0.6cm,y=1.2cm,->,shorten >=1pt]
\tikzstyle{mynode}=[draw=black,fill=white,circle,minimum size=3,scale=0.6]
\def\L{16}
\def\pos{\L+0.8}
\def\z{0.18}
\foreach \N [count=\lay,remember={\N as \Nprev (initially 0);}] in {1,1,2,3,8,9,10,10}{ % loop over layers
\draw[fill=blue!20] (\N-2*\z,\z-\lay) -- (\N-2*\z,-\z-\lay) -- (\L+2*\z,-\z-\lay) -- (\L+2*\z,\z-\lay) -- cycle;
\foreach \x [evaluate={\y=-\lay; \prev=int(\lay-1);\xprev=int(\x-\N+\Nprev);}] in {\N,...,\L}{ % loop over nodes
  \node[mynode] (N\lay-\x) at (\x,\y) {};
  \ifthenelse{\lay>1}
    {\foreach \j in {\xprev,...,\x}
      \draw (N\prev-\j) -- (N\lay-\x);}{}
}
}
\node[right] at (\pos,-1) {$\mZ$};
\node[right] at (\pos,-2) {$\m{H}^{(0)}$};
\node[right] at (\pos,-4) {$\m{H}^{(1)}$};
\node[right] at (\pos,-5) {$\m{H}^{(2)}$};
\node[right] at (\pos,-7) {$\m{H}^{(3)}$};
\node[right] at (\pos,-8) {$\mY$};
\node[right] at (\pos,-4.5) {$\text{attn}_{\texttt{c}}$};
\node[right] at (\pos,-1.5) {$\text{conv}_{\texttt{c}}$};
\node[right] at (\pos,-2.5) {$\text{conv}_{\texttt{c}}$};
\node[right] at (\pos,-3.5) {$\text{conv}_{\texttt{c}}$};
\node[right] at (\pos,-5.5) {$\text{conv}_{\texttt{c}}$};
\node[right] at (\pos,-6.5) {$\text{conv}_{\texttt{c}}$};
\node[right] at (\pos,-7.5) {$\text{conv}_{\texttt{c}}$};
\end{tikzpicture}
\caption{Illustration of the generator of \tagan{} with $n_k=2$ and $L_{2}^{G}=L_{2}^{G}=1$.}
\label{fig:tagan_illu}
\end{figure}
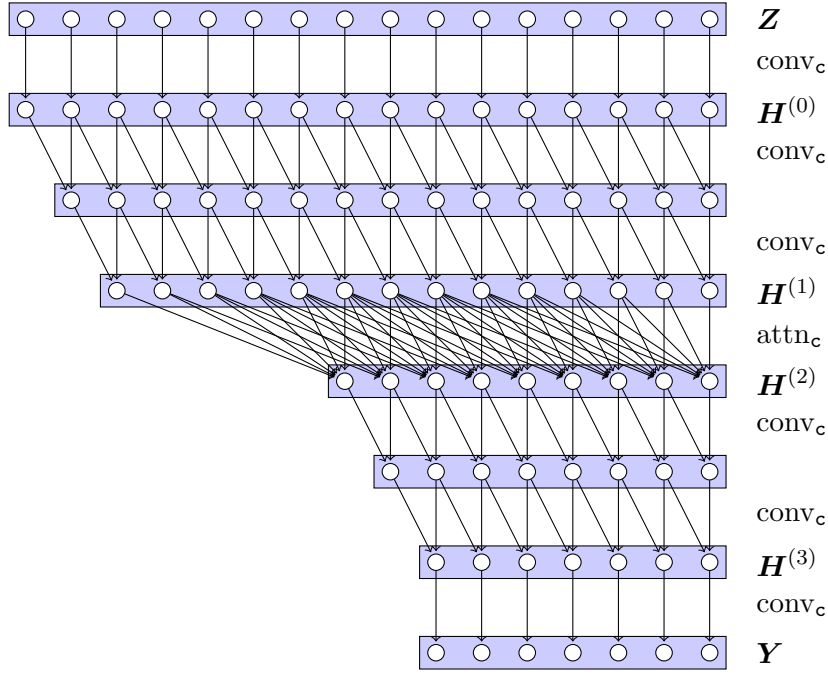

Suppose the input noise of the generator is $\mZ\in\R^{(l+f-1)\times d_n}$ and the output sample is $\mY\in\R^{l\times d}$. With the notations of the network layers introduced in Section \ref{sec:layer}, the generator can be written as follows
\begin{align*}
	\m{H}^{(0)} &= \convc{1}{d_h} \left(\mZ\right)\\
	\m{H}^{(j)} &= \convc{n_k}{d_h} \circ h_G \circ \convc{n_k}{d_h} \circ h_G \left(\m{H}^{(j-1)}\right),\forall 1\leq j\leq L_{1}^{G}\\
	\m{H}^{(L_{1}^{G}+1)} &= \attnc{n_{a,G}}{n_h}{f-2(L_1+L_2)(n_k-1)} \left(\m{H}^{(L_{1}^{G})}\right)\\
	\m{H}^{(j)} &= \convc{n_k}{d_h} \circ h_G \circ \convc{n_k}{d_h} \circ h_G \left(\m{H}^{(j-1)}\right),\forall L_{1}^{G}+2\leq j\leq L_{2}^{G}+1\\
	\mY &= \convc{1}{d} \circ h_G\left(\m{H}^{(L_{2}^{G}+1)}\right)
\end{align*}
where $\circ$ means composition of the layers. The attention layer is embedded in the middle of the network to extend the receptive field and the convolutional layers are responsible for learning the local characteristics. The network structure of the generator of \tagan{} is illustrated in Figure \ref{fig:tagan_illu}. 

Let $\mY\in\R^{l\times d}$ denote either the real data or the fake data from the generator and $D(\mY;\theta_D)$ be the output of the discriminator. For the notation simplicity, we let
\begin{align*}
\zeta_1^{(j)}=\convr{n_k}{\min(2^{j-1}d_s,d_m)}{1}\\
\zeta_2^{(j)}=\convr{n_k}{\min(2^{j-1}d_s,d_m)}{2}.
\end{align*} 
Then the discriminator can be written as follows:
\begin{align*}
	\m{U}^{(0)} =&\, \mY\\
	\m{U}^{(j)} =&\, \zeta_2^{(j)} \circ h_D \circ \zeta_1^{(j)} \circ h_D\left(\m{U}^{(j-1)}\right),\forall 1\leq j\leq L_{1}^{D}\\
	\m{U}^{(L_{1}^{D}+1)} =&\, \attnr{n_{a,D}}{n_h} \left(\m{U}^{(L_{1}^{D})}\right)\\
	\m{U}^{(j)} =&\, \zeta_2^{(j-1)} \circ h_D \circ \zeta_1^{(j-1)} \circ h_D\left(\m{U}^{(j-1)}\right),\forall L_{1}^{D}+2\leq j\leq L_{2}^{D}+1\\
	D(\mY;\theta_D) =&\, \sum_{i_1=1}^{n_l}\sum_{i_2=1}^{\min(2^{L_{1}^{D}+L_{2}^{D}-1}d_s,d_m)}h_D\left({U}_{i_1,i_2}^{(L_{2}^{D}+1)}\right)w_{i_2}
\end{align*} 
where $\m{w}\in\R^{\min(2^{L_{1}^{D}+L_{2}^{D}-1}d_s,d_m)}$ is a trainable weight used before the output in the discriminator. The discriminator is the classical architecture of the convolutional network which shrinks the length but increases the channel of hidden layers. 

We apply batch normalization \cite{ioffe_batch_2015} before activation functions, spectrum normalization \cite{miyato2018spectral} to the layer weights, and residual connections \cite{he_2016_CVPR} and skip connections \cite{wiese_quant_2020} to the generator and discriminator in \tagan{} during training to stabilize the statistics of generated samples and improve the performance. Batch normalization normalizes the output of the layers, and the spectrum normalization limits the spectral norm of the weight parameters, both of which reduce extreme values in the network. Residual connections and skip connections accelerate training when the networks become deep.

\subsection{Temporal transformer \gan{}}
The temporal transformer \gan{} (\ttgan{}) is composed of two transformer networks as its generator and discriminator. Similar models can be found in \cite{jiang_gan_2021,hudson_gan_2021}. A transformer consists of several attention layers with each layer followed by a two-layer MLP. We use the causal attention layers in the generator and the sparse attention layer in the discriminator. Each causal attention layer has a flexible RFS. The hyper-parameters of \ttgan{} are listed in Table \ref{tab:ttgan_paras}.

\begin{table}[h]
\centering
	\begin{tabular}{rl}\hline
	hyper-parameter & meaning \\ \hline
	$l$ & data length \\
	$f$ & receptive field size \\
	$d_n$ & noise channel \\
	$d$ & data channel \\
	$d_h$ & hidden channel \\
	$n_h$ & number of heads in attention\\
	$n_a$ & attention hidden size\\
	$L$ & number of attention layers\\
	$\{f_j\}_{j=1}^{L}$ & the RFS of attentions in the generator\\
	$n_m$ & hidden size in the multi-layer perceptron\\
	$h(\cdot)$ & activation function\\
	\hline
	\end{tabular}
	\caption{Hyper-parameters in \ttgan{}.}
	\label{tab:ttgan_paras}
\end{table}

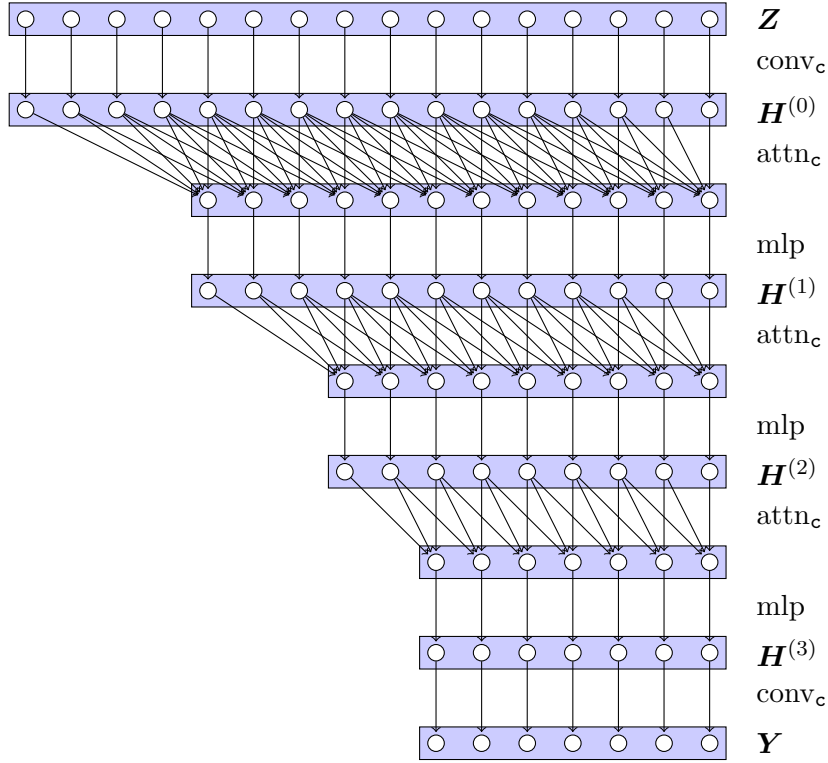
\begin{figure}[h]
\centering
\begin{tikzpicture}[x=0.6cm,y=1.2cm,->,shorten >=1pt]
\tikzstyle{mynode}=[draw=black,fill=white,circle,minimum size=3,scale=0.6]
\def\L{16}
\def\pos{\L+.8}
\def\z{0.18}
\foreach \N [count=\lay,remember={\N as \Nprev (initially 0);}] in {1,1,5,5,8,8,10,10,10}{ % loop over layers
\draw[fill=blue!20] (\N-2*\z,\z-\lay) -- (\N-2*\z,-\z-\lay) -- (\L+2*\z,-\z-\lay) -- (\L+2*\z,\z-\lay) -- cycle;
\foreach \x [evaluate={\y=-\lay; \prev=int(\lay-1);\xprev=int(\x-\N+\Nprev);}] in {\N,...,\L}{ % loop over nodes
  \node[mynode] (N\lay-\x) at (\x,\y) {};
  \ifthenelse{\lay>1}
    {\foreach \j in {\xprev,...,\x}
      \draw (N\prev-\j) -- (N\lay-\x);}{}
}
}
\node[right] at (\pos,-1) {$\mZ$};
\node[right] at (\pos,-2) {$\m{H}^{(0)}$};
\node[right] at (\pos,-4) {$\m{H}^{(1)}$};
\node[right] at (\pos,-6) {$\m{H}^{(2)}$};
\node[right] at (\pos,-8) {$\m{H}^{(3)}$};
\node[right] at (\pos,-9) {$\mY$};
\node[right] at (\pos,-2.5) {$\text{attn}_{\texttt{c}}$};
\node[right] at (\pos,-4.5) {$\text{attn}_{\texttt{c}}$};
\node[right] at (\pos,-6.5) {$\text{attn}_{\texttt{c}}$};
\node[right] at (\pos,-3.5) {$\text{mlp}$};
\node[right] at (\pos,-5.5) {$\text{mlp}$};
\node[right] at (\pos,-7.5) {$\text{mlp}$};
\node[right] at (\pos,-1.5) {$\text{conv}_{\texttt{c}}$};
\node[right] at (\pos,-8.5) {$\text{conv}_{\texttt{c}}$};
\end{tikzpicture}
\caption{Illustration of the generator of \ttgan{} with $L=3$ and $(f_1,f_2,f_3)=(5,4,3)$.}
\label{fig:ttgan_illu}
\end{figure}

Suppose the input noise of the generator is $\mZ\in\R^{(l+f-1)\times d_n}$ and the output sample is $\mY\in\R^{l\times d}$. With the notations of the network layers introduced in Section \ref{sec:layer}, the generator can be written as follows
\begin{align*}
	\m{H}^{(0)} &= \convc{1}{d_h} \left(\mZ\right)\\
	\m{H}^{(j)} &= \mlp{n_m}{h} \circ \attnc{n_a}{n_h}{f_j} \left(\m{H}^{(j-1)}\right),\forall 1\leq j\leq L\\
	\mY &= \convc{1}{d} \left(\m{H}^{(L)}\right)
\end{align*}
where $\circ$ means composition of the layers. The RFS of each attention layer needs to satisfy the equation $f-1=\sum_{j=1}^{L}(f_j-1)$. This is because the length shrinkage of each attention layer is equal to RFS$-1$ and the total length shrinkage is equal to the sum of the shrinkage of all attention layers. The network structure of the generator of \ttgan{} is illustrated in Figure \ref{fig:ttgan_illu}. At a first glance at Figures \ref{fig:tagan_illu} and \ref{fig:ttgan_illu}, it seems as if there is no difference between attention layers and convolutional layers. However, one should note that the formula of attention layers is different from that of convolutional layers and the RFS of attention layers can be much larger.

Let $\mY\in\R^{l\times d}$ denote either the real data or the fake data from the generator and $D(\mY;\theta_D)$ be the output of the discriminator. The discriminator can be written as follows.  
\begin{align*}
	\m{U}^{(0)} &= \convr{1}{d_h}{1} \left(\mY\right)\\
	\m{U}^{(j)} &= \mlp{n_m}{h} \circ \attns{n_a}{n_h} \left(\m{U}^{(j-1)}\right),\forall 1\leq j\leq L\\
	D(\mY;\theta_D) &= \sum_{i_1=1}^{l}\sum_{i_2=1}^{n_h}{U}_{i_1,i_2}^{(L)}W_{i_1,i_2}
\end{align*} 
where $\m{W}\in\R^{l\times n_h}$ is a trainable weight used before the output in the discriminator. The discriminator is the classical architecture of the transformer encoder but with sparse attention.

We apply batch normalization \cite{ioffe_batch_2015} before each attention layer and MLP block, spectrum normalization \cite{miyato2018spectral} to the layer weights, residual connections \cite{he_2016_CVPR} and skip connections \cite{wiese_quant_2020} to the generator and discriminator in \ttgan{} during training to stabilize the statistics of generated samples and improve the performance.

\section{Simulation of the S\&P 500 index}\label{sec:index_simulation}
In this section, we show the numerical results of the proposed networks for the S\&P 500 index simulation. 

\subsection{Stylized facts and metrics}
It has been summarized in \cite{cont_2001} that the returns of the S\&P 500 index and also the equities admit the following characteristics called \textit{stylized facts}:
\begin{itemize}
	\item Asset returns shows heavier tails than the normal distribution.
	\item Escalator up and elevator down: large drawdowns but not equally large upward movements are observed.
	\item Autocorrelations of (daily) asset returns are often insignificant.
	\item Volatility clustering: volatility displays a positive autocorrelation.
	\item The autocorrelation function (ACF) of absolute returns decays slowly as a function of the time lag.
	\item Leverage effect: volatility of an asset is negatively correlated with the returns of that asset.
\end{itemize}

Suppose we have a sequence of historical prices $p_{0:T_x}=\{p_t\}_{t=0}^{T_x}$. We take the log returns to be the real data, i.e. $x_{1:T_x}=\{x_t\}_{t=1}^{T_x}$ where $x_t =\ln(p_t/p_{t-1})$. Then we sample $N$ sequences of returns from the \gan{} and denote them as $\{y_{1:T}^{(i)}\}_{i=1}^{N}$. The evaluation metrics are listed as follows:
\begin{itemize}
\item The Wasserstein-1 distance of daily and multi-day returns. Let $F_{\tau}^h(x)$ denote the empirical CDF of the historical $\tau$-day returns 
$$\left\{\sum_{j=0}^{\tau-1}x_{t+j}:1\leq t\leq T_x-\tau+1\right\}$$ 
and $F_{\tau}^g(x)$ the empirical CDF of the generated $\tau$-day returns 
$$\left\{\sum_{j=0}^{\tau-1}y_{t+j}^{(i)}:1\leq i\leq N,1\leq t\leq T_x-\tau+1\right\}.$$
The Wasserstein-1 distance is given by 
\begin{align}
W_1^{(\tau)}(x_{1:T_x},\{y_{1:T}^{(i)}\}_{i=1}^{N})=\int_{\R}\vert F_{\tau}^g(x)-F_{\tau}^h(x)\vert dx.\label{eq:w1_dist}
\end{align}
We will calculate the Wasserstein-1 distance of 1-, 5-, 20-, 100- and 200-day returns.

\item High order moment scores: skewness and kurtosis. We calculate \begin{align*}
	\left| \text{skew}(x_{1:T_x}) - \frac{1}{N}\sum_{1\leq i\leq N}\text{skew}\left(y^{(i)}_{1:T}\right)\right|
\end{align*} and \begin{align*}
	\left| \text{kurt}(x_{1:T_x}) - \frac{1}{N}\sum_{1\leq i\leq N}\text{kurt}\left(y^{(i)}_{1:T}\right)\right| 
\end{align*} 
as the scores of skewness and kurtosis.

\item Correlation scores. We look at the following four scores:
\begin{itemize}
\item Autocorrelation of returns: ACF$_{\tau}(x_{1:T_x})=\text{corr}(x_t,x_{t+\tau})$
\item Autocorrelation of absolute returns: ACF$^{(\text{abs})}_{\tau}(x_{1:T_x})=\text{corr}(\vert x_t\vert,\vert x_{t+\tau}\vert)$
\item Autocorrelation of squared returns: ACF$^{(\text{sq})}_{\tau}(x_{1:T_x})=\text{corr}(x_t^2,x_{t+\tau}^2)$
\item Leverage effect: Lev$_{\tau}(x_{1:T_x})=\text{corr}(x_t,x_{t+\tau}^2)$
\end{itemize}
Each score is calculated for lag $1\leq \tau\leq \delta$. Then we calculate \begin{align*}
	\sqrt{\sum_{1\leq \tau\leq \delta}\left( \text{score}_{\tau}(x_{1:T_x}) - \frac{1}{N}\sum_{1\leq i\leq N}\text{score}_{\tau}\left(y^{(i)}_{1:T}\right)\right)^2}
\end{align*}
where score stands for ACF, ACF$^{(\text{abs})}$, ACF$^{(\text{sq})}$ and Lev.
\end{itemize}

Those metrics do not participate in the loss functions of the \gans{} during training, so they are suitable to evaluate the samples generated from the \gans{}.

\subsection{Training}
After we have the real data $x_{1:T_x}$, we apply a rolling window of length $l$ to get the real dataset $\{{x}_{t:t+l-1}\}_{t=1}^{T_x-l+1}$ for training. 

One important stylized fact of the asset returns is that tails of their distributions are heavier than that of the normal distribution. However, we usually use the normal distribution as the random noise input of \gans{}. Thus it is a question whether \gans{} are able to generate heavy-tailed distributions given normal noise. In \cite{wiese_quant_2020}, the authors use the inverse Lambert transform to make the returns closer to the normal distribution such that \gans{} do not need to generate heavy tails. But in our experiments, we still use the original returns as the training data. We would like to show that the proposed \gans{} can learn to generate heavy tails by themselves even if they are given normal noise input.

We also test the case of adding the cumulative sum of the returns as additional channels to the input of the discriminator. The output from the generator is a sequence ${y}_{1:l}$. Then the augmented input to the discriminator is $\tilde\mY\in\R^{l\times 2}$ where $$\tilde{Y}_{t,j}=\begin{cases}
	{y}_{t,j}, &\text{if}\,j=1\\
	\sum_{i_1=1}^{t}{y}_{i_1,j}, &\text{if}\,j=2.\\
\end{cases}$$
In this way, the input of the discriminator includes not only the returns, but also the log-prices starting from 0. The additional cumulative sum feature is added so that the discriminator can observe the returns over large intervals by taking the difference of the log-prices at two time points instead of taking the cumulative sum of the returns over long intervals.

In the rest of this section, we compare the performance of the proposed \tagan{} and \ttgan{} with \quantgan{} \cite{wiese_quant_2020}, which has shown good results for the S\&P 500 index simulation using the temporal convolutional network \cite{vandenoord16_ssw}. The data channel is $d=1$. The data length is $l=128$ for \tagan{} and \ttgan{} while $l=127$ for \quantgan{}. The RFS is $f=127$ for each \gan{}. The number of layers in the generator is $L_1^G=L_2^G=3$ for \tagan{} and $L=5$ for \ttgan{}. The number of hidden channels is 64 in \tagan{} and \ttgan{} and is 80 in \quantgan{}. We calculate $M=512$ simulated paths of length $T=2560$ for evaluation. In the correlation scores, we let $\delta=250$ in accordance with \cite{wiese_quant_2020}. The loss for training \quantgan{} is the loss of the original \gan{} in Equation \eqref{eq:loss_gan}, as used in their paper. We use the loss functions of the \wgangp{} in Equation \eqref{eq:loss_wgan_gp} to train \tagan{} and \ttgan{}.  

\subsection{Simulation of the medium kurtosis data}
To make a fair comparison with \quantgan{}, we use the same data in the paper of \quantgan{} \cite{wiese_quant_2020}, which is the S\&P 500 index daily data from May 1, 2009 to Nov 30, 2018 with $T_x=2414$.  The skewness of the data is -0.4667 and the kurtosis is 4.0648.

We test \tagan{}, \ttgan{} and \quantgan{} with and without the additional cumulative sum feature. We only present the cases of good performance since not every case works. The selected results of the three \gans{} without the additional cumulative sum feature, as well as \ttgan{} with the additional cumulative sum feature, are shown in Table \ref{tab:sp500_mid_kurt}. Here is the summary of results:
\begin{itemize}
	\item The performance of the four candidates in Table \ref{tab:sp500_mid_kurt} are close to each other and the difference is not significant.
	\item The cumulative sum feature only improves the performance of \ttgan{}. This means the transformer is more suitable to process features of different scales than the convolutional network. With the help of the cumulative sum feature, \ttgan{} reduces the Wasserstein-1 distance score of 200-day returns, which agrees with the purpose of the additional cumulative sum feature.
\end{itemize}

\begin{table}[h]
\centering
	\begin{tabular}{ccccc}\hline
	scores & \tagan{} & \begin{tabular}{c}\ttgan{}\\(w/o cumsum)\end{tabular} & \begin{tabular}{c}\ttgan{}\\(w/ cumsum)\end{tabular} & \quantgan{} \\\hline
	$W_1^{(1)}$ & 4.569e-04 & 2.143e-04 & 3.319e-04 & 2.940e-04\\
	$W_1^{(5)}$ & 9.764e-04 & 4.803e-04 & 7.367e-04 & 6.999e-04\\
	$W_1^{(20)}$ & 2.677e-03 & 1.574e-03 & 2.234e-03 & 1.800e-03\\
	$W_1^{(100)}$ & 3.363e-03 & 4.338e-03 & 3.311e-03 & 4.952e-03\\
	$W_1^{(200)}$ & 1.016e-02 & 1.128e-02 & 7.281e-03 & 1.377e-02\\
	skewness & 5.284e-02 & 1.110e-01 & 1.752e-01 & 2.014e-01\\
	kurtosis & 5.248e-01 & 3.363e-01 & 1.237e-01 & 3.096e-01\\
	ACF & 3.450e-01 & 3.609e-01 & 3.628e-01 & 3.420e-01\\
	ACF$^{(\text{abs})}$ & 3.799e-01 & 3.727e-01 & 3.552e-01 & 3.742e-01\\
	ACF$^{(\text{sq})}$ & 3.300e-01 & 3.274e-01 & 3.238e-01 & 3.301e-01\\
	Lev & 3.248e-01 & 3.368e-01 & 3.376e-01 & 3.305e-01\\
	\hline
	\end{tabular}
	\caption{Scores of the S\&P 500 index simulation given the medium kurtosis data from May 1, 2009 to Nov 30, 2018.}
	\label{tab:sp500_mid_kurt}
\end{table}

\subsection{Simulation of the high kurtosis data}
To further test the ability of the \gans{} to generate data with high (negative) skewness and high kurtosis, we also use the S\&P 500 index daily data from May 1, 2009 to Dec 31, 2020 as the training data, which includes the drawdowns in 2020. The size of the dataset is $T_x=2938$, the skewness is -0.8132 and the kurtosis is 15.1333.  

We test \tagan{}, \ttgan{} and \quantgan{} for the dataset. For \tagan{} and \quantgan{}, no cumulative sum is used, while for \ttgan{}, we always use the cumulative sum feature. We also test \ttgan{} using batch normalization by default and its variant where we replace batch normalization with layer normalization \cite{ba2016layer}. Layer normalization normalizes the input values across the features, while batch normalization normalizes the input values across the batch dimension. The results are summarized in Table \ref{tab:sp500_high_kurt}. The results of \tagan{}, \ttgan{} with batch normalization and \quantgan{} are further illustrated in Figure \ref{fig:tagan_index}, \ref{fig:ttgan_index} and \ref{fig:quantgan_index}. Here are the summary of the results:
\begin{itemize}
	\item All the \gans{} perform well in fitting the distribution.
	\item Although layer normalization is more often used in the transformer architecture, we found that the layer normalization transformer fails to generate samples with high kurtosis in our tests.
	\item The convolution-based \gans{}, \tagan{} and \quantgan{}, are very sensitive to the autocorrelation curves, while \ttgan{} tends to smooth the autocorrelation curves. The fluctuations in the autocorrelation curves are likely to be caused by randomness in the market. The convolution-based \gans{} are preferred if we need to replicate the realization of randomness, while the transformer-based \ttgan{} is more suitable if we would like to filter out the randomness.
\end{itemize}

\begin{table}[h]
\centering
\begin{tabular}{ccccc}\hline
	scores & \tagan{} & \ttgan{} (BN) & \ttgan{} (LN) & \quantgan{}\\\hline
	$W_1^{(1)}$ & 4.823e-04 & 4.907e-04 & 2.431e-04 & 2.605e-04\\
	$W_1^{(5)}$ & 1.097e-03 & 1.525e-03 & 7.800e-04 & 9.530e-04\\
	$W_1^{(20)}$ & 2.844e-03 & 4.963e-03 & 1.804e-03 & 2.840e-03\\
	$W_1^{(100)}$ & 5.542e-03 & 8.432e-03 & 1.265e-02 & 6.347e-03\\
	$W_1^{(200)}$ & 2.050e-02 & 1.774e-02 & 3.033e-02 & 1.797e-02\\
	skewness & 2.539e-01 & 4.883e-02 & 1.663e-01 & 3.870e-02\\
	kurtosis & 2.173e-01 & 2.121e-01 & 4.591e+00 & 5.674e-01\\
	ACF & 3.323e-01 & 4.067e-01 & 4.273e-01 & 3.437e-01\\
	ACF$^{(\text{abs})}$ & 3.792e-01 & 3.465e-01 & 3.740e-01 & 3.647e-01\\
	ACF$^{(\text{sq})}$ & 2.409e-01 & 2.496e-01 & 3.175e-01 & 2.415e-01\\
	Lev & 2.300e-01 & 2.957e-01 & 2.945e-01 & 2.319e-01\\
	\hline
\end{tabular}
\caption{Scores of the S\&P 500 index simulation given the high kurtosis data from May 1, 2009 to Dec 31, 2020.}
\label{tab:sp500_high_kurt}
\end{table}

\section{Simulation of the option surface}\label{sec:option_simulation}
In this section, we show the numerical results of the proposed networks for the option surface simulation.

\subsection{Formulation}
Suppose we have $N_K$ relative strikes
$$\mathcal{K}=\{K_1,K_1+\Delta K,\dots,K_1+(N_K-1)\Delta K\}$$
and $N_M$ maturities
$$\mathcal{M}=\{M_1,M_2,\dots,M_{N_M}\}.$$
Let $d=N_M\times N_K$. The real data is $\{\vx_t\in\R^{d}\}_{t=1}^{T_x}$ with the elements $\vx_t=(x_{t,j})_{j=1}^{d}$, where $x_{t,(j_1-1)N_M+j_2}=\ln{\sigma}_{t,(j_1-1)N_M+j_2}$ is the log-volatility at time $t$ with the relative strike $K_{j_2}=K_1+(j_2-1)\Delta K$ and the maturity $M_{j_1}$.

\subsection{Training}
Having the real data $\vx_{1:T_x}=\{\vx_t\}_{t=1}^{T_x}$, we apply a rolling window of length $l$ to get the dataset $\{{\vx}_{t:t+l-1}\}_{t=1}^{T_x-l+1}$ for training. The output $\{\hat{\vy}_{1:T}^{(i)}\}_{i=1}^{N}$ from the \gan{} generator is not guaranteed to be arbitrage-free. We apply the method in Appendix \ref{app:no_arbitrage} to detect and remove arbitrage to obtain the arbitrage-free surface $\{{\vy}_{1:T}^{(i)}\}_{i=1}^{N}$. In \cite{wiese2019deep}, the authors use the discrete local volatilities \cite{buehler2017discrete} to replace the implied volatilities when generating arbitrage-free option surfaces. The proposed networks are compatible with discrete local volatilities, but we still expect them to generate the implied volatilities and examine to what extent the outputs from the \gans{} violate the no-arbitrage condition.

\begin{figure}
\centering
\tikzstyle{block} = [rectangle, draw, 
    text width=7em, text centered, rounded corners, minimum height=2em]
\tikzstyle{line} = [draw, -latex']
\begin{tikzpicture}[node distance = 1.5cm, auto]
    \node [block] (b0) at (0,0) {$\vx_{1:T_x}$};
    \node [block] (b1) at (5cm,0) {$\tilde\vx_{1:T_x}$};
    \node [block] (b2) at (10cm,0) {$\{\tilde{\vx}_{t:t+l-1}\}_{t=1}^{T_x-l+1}$};
    \node [block] (b3) at (10cm,-2cm) {$\{\tilde{\vy}_{1:T}^{(i)}\}_{i=1}^{N}$};
    \node [block] (b4) at (5cm,-2cm) {$\{\hat{\vy}_{1:T}^{(i)}\}_{i=1}^{N}$};
    \node [block] (b5) at (0cm,-2cm) {$\{{\vy}_{1:T}^{(i)}\}_{i=1}^{N}$};
    
    \path [line] (b0) -- node[above] {PCA} (b1) ;
	\path [line] (b1) -- node[above] {rolling} node[below] {window} (b2);
    \path [line] (b2) -- node[left] {simulate} (b3);
    \path [line] (b3) -- node[above] {reverse} node[below] {PCA} (b4);
    \path [line] (b4) -- node[above] {remove} node[below] {arbitrage} (b5);
\end{tikzpicture}
\caption{Pipeline of \gans{} using PCA.}
\label{fig:pca_flow_chart}
\end{figure}
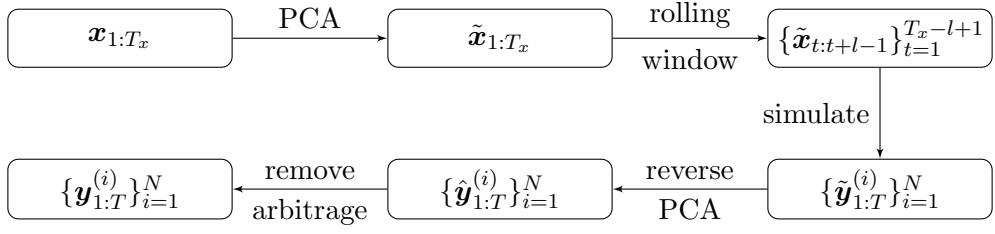

The option volatility data is a high-dimensional data with high cross-correlation, so we could use principal component analysis (PCA) to reduce dimensionality. We perform PCA on the original data $\vx_{1:T_x}$ and get the first $\tilde{d}$ principal components $\{\tilde{\vx}_t\in\R^{\tilde{d}}\}_{t=1}^{T_x}$. To be more specific, suppose the real data matrix $\mX\in\R^{T_x\times d}$ is $\mX=\left(\vx_1,\vx_2,\dots,\vx_{T_x}\right)^{\top}$. We get its singular value decomposition (SVD) as $\mX=\m{U}\m{D}\m{V}^{\top}$, and then take the first $\tilde{d}$ columns of $\m{U}$ to be the principal components, i.e. $\m{U}_{\cdot,1:\tilde{d}}=\left(\tilde{\vx}_1,\tilde{\vx}_2,\dots,\tilde{\vx}_{T_x}\right)^{\top}$. Next, we apply a rolling window of length $l$ to get the real dataset $\{\tilde{\vx}_{t:t+l-1}\}_{t=1}^{T_x-l+1}$ for training. The \gan{} generator is responsible for generating the first $\tilde{d}$ principal components $\{\tilde{\vy}_{1:T}^{(i)}\}_{i=1}^{N}$, and they are used to recover the log-volatility surfaces $\{\hat{\vy}_{1:T}^{(i)}\}_{i=1}^{N}$ through reverse PCA, where $\hat{\vy}_t^{(i)}=\m{V}_{\cdot,1:\tilde{d}}\,\m{D}_{1:\tilde{d},1:\tilde{d}}\,\tilde{\vy}_t^{(i)}$. Finally we apply the method in Appendix \ref{app:no_arbitrage} to get the arbitrage-free surfaces $\{{\vy}_{1:T}^{(i)}\}_{i=1}^{N}$. This process is summarized in Figure \ref{fig:pca_flow_chart}.

Since we find it is helpful to include both returns and log-prices in the S\&P 500 index simulation, we think it could also be helpful when the differences of the log-volatilities (called log-volatility returns hereafter) are used as the additional feature. The output from the generator is a sequence $\hat\vy_{1:l}$. Then the augmented input to the discriminator is ${\bar{\mY}}\in\R^{l\times 2d}$ where $${\bar{Y}}_{t,j}=\begin{cases}
	\hat{y}_{t,j}, &\text{if}\,1\leq j\leq d\\
	\hat{y}_{t,j-d}-\hat{y}_{t-1,j-d}, &\text{if}\,2\leq t\leq l\,\text{and}\,d+1\leq j\leq 2d\\
	0, &\text{if}\,t=1\,\text{and}\,d+1\leq j\leq 2d.\\
\end{cases}$$
In this way, the input of the discriminator includes both the log-volatilities and the log-volatility returns.

To summarize, we have three choices to train the \gans{}: 
\begin{itemize}
	\item Use the log-volatility surfaces as the real data. The generators simulate the log-volatility surfaces.
	\item Use the principal components of log-volatility surfaces as the real data. The generators simulate the principal components.
	\item Use the log-volatility surfaces and their returns as the real data. The generators simulate the log-volatility surfaces.
\end{itemize}

\subsection{Stylized facts and metrics}
\begin{figure}[h]
    \centering
 	\includegraphics[width = 0.6\textwidth]{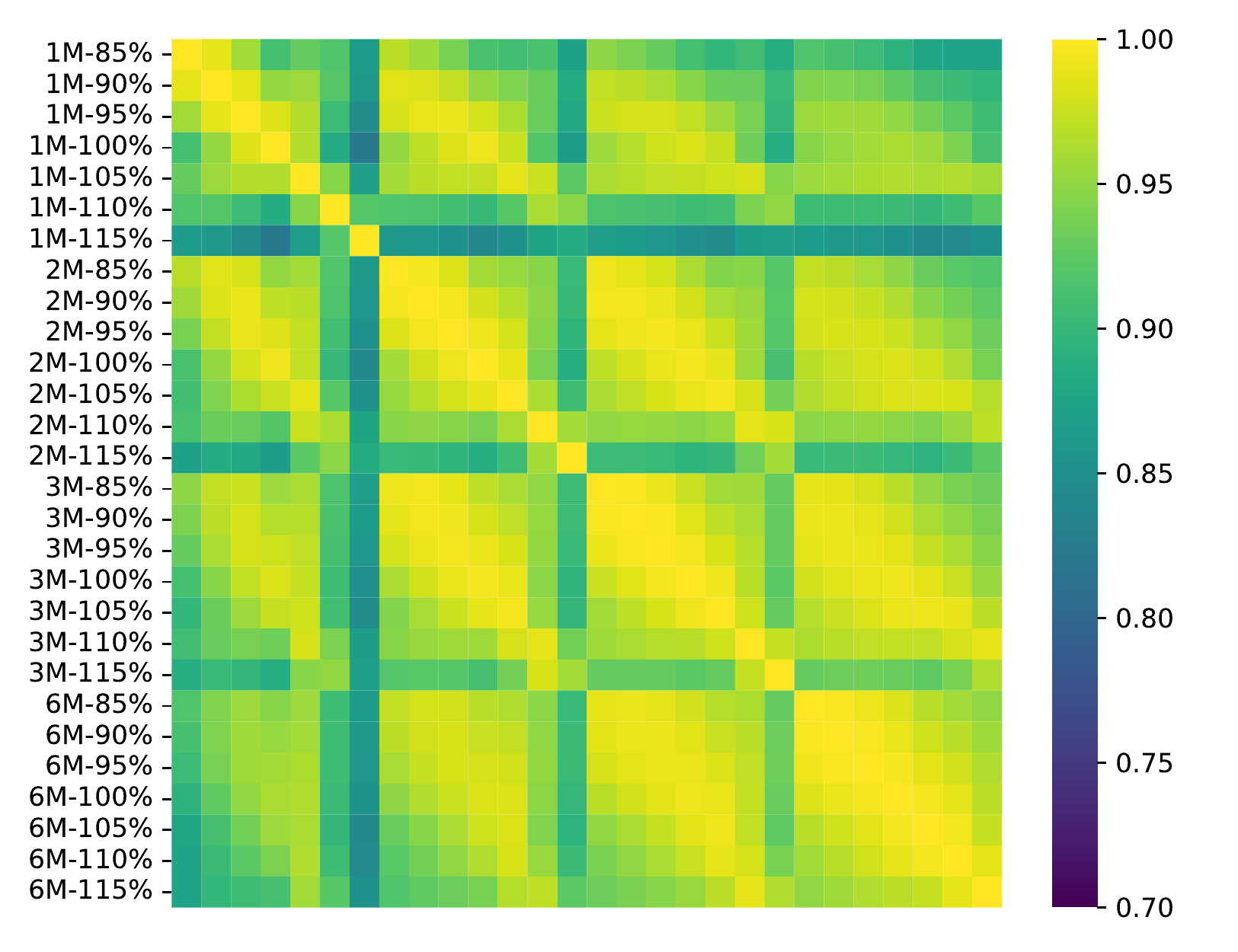}
 	\caption{Cross-correlation matrix of the log-volatilities of the S\&P 500 index options. The label means `maturity - relative strike'.}
    \label{fig:vol_corr}
\end{figure}

Here are some stylized facts of the option surface summarized in \cite{wiese2021multi}.
\begin{itemize}
	\item The volatility smile. Deep in-the-money and out-of-the-money volatility are generally higher than at-the-money volatility.
	\item Volatilities have high serial autocorrelation.
	\item Volatilities show high cross-correlation. The correlation matrix of the log-volatilities of different relative strikes and maturities in the S\&P 500 index option data in Section \ref{subsec:option_data} is shown in Figure \ref{fig:vol_corr}. Higher cross-correlation is observed for proximate relative strikes and maturities. Volatilities of longer maturities have higher cross-correlation than volatilities of shorter maturities.
\end{itemize}

Based on the stylized facts, the evaluation metrics are listed as follows:
\begin{itemize}
\item The Wasserstein-1 distance of distribution of volatilities \begin{align*}
\frac{1}{d} \sum_{j=1}^{d}W_1^{(1)}\left(\{x_{t,j}\}_{t=1}^{T_x},\{\{y_{t,j}^{(i)}\}_{t=1}^{T}\}_{i=1}^N\right)
\end{align*}
where the Wasserstein-1 distance is already defined in Equation \eqref{eq:w1_dist}.

\item High order moment scores of skewness
\begin{align*}
	\frac{1}{d} \sum_{j=1}^{d}\left| \text{skew}(x_{1:T_x,j}) - \frac{1}{N}\sum_{1\leq i\leq N}\text{skew}\left(y^{(i)}_{1:T,j}\right)\right|
\end{align*} and kurtosis \begin{align*}
	\frac{1}{d} \sum_{j=1}^{d}\left| \text{kurt}(x_{1:T_x,j}) - \frac{1}{N}\sum_{1\leq i\leq N}\text{kurt}\left(y^{(i)}_{1:T,j}\right)\right| .
\end{align*} 

\item Autocorrelation score of the series and returns. Define $$\text{ACF}_{\tau}^{(\text{r})}(x_{1:T_x})=\text{ACF}_{\tau}(x_{2:T_x}-x_{1:T_x-1}).$$ We calculate
\begin{align*}
	\frac{1}{d} \sum_{j=1}^{d}\sqrt{\sum_{1\leq \tau\leq \delta}\left( \text{score}_{\tau}(x_{1:T_x,j}) - \frac{1}{N}\sum_{1\leq i\leq N}\text{score}_{\tau}\left(y^{(i)}_{1:T,j}\right)\right)^2}
\end{align*}
where score stands for ACF and ACF$^{(\text{r})}$.

\item Cross-correlation score. Let $\Sigma_{\vx}\in \R^{d\times d}$ be the cross-correlation matrix of $\{\vx_{t},1\leq t\leq T_x\}$ and $\Sigma_{\vy}\in \R^{d\times d}$ be the cross-correlation matrix of $\{\vy_t^{(i)},1\leq t\leq T_x,1\leq i\leq N\}$. Then the score is defined to the Frobenius norm of the difference $\Vert\Sigma_{\vx}-\Sigma_{\vy}\Vert_{F}$.
\item Arbitrage rate. The score is calculated as the percentage of the outputs from the \gans{} $\hat\vy^{(i)}_{1:T}$ that violate the no-arbitrage condition $$\#\{(i,t):\hat{\vy}^{(i)}_{t}\,\text{that violates the no-arbitrage condition},1\leq i\leq N,1\leq t\leq T\}/(NT).$$ This is a score that shows how well the \gan{} can learn the no-arbitrage condition.

\end{itemize}

\subsection{Data and results}\label{subsec:option_data}
We use the daily data of the S\&P 500 index options from Jan 02, 2009 to Oct 30, 2020 as the real data. The maturities are $$\mathcal{M}=\{\text{1-month, 2-month, 3-month, 6-month}\}$$ and the relative strikes are
$$\mathcal{K}=\{85\%,90\%,95\%,100\%,105\%,110\%,115\%\}.$$
The data channel is $d=28$ and the sequence length is $T_x=2979$.

We compare the performance of the proposed \tagan{} and \ttgan{} with \quantgan{} \cite{wiese_quant_2020} and try both PCA and the additional log-volatility return feature. The data length is $l=128$ for \tagan{} and \ttgan{} while $l=127$ for \quantgan{}. The RFS is $f=383$ for each \gan{}. We let $\tilde{d}=10$ for PCA. The number of layers in the generator is $L_1^G=L_2^G=3$ for \tagan{} and $L=5$ for \ttgan{}. The number of hidden channels is 64 in \tagan{} and \ttgan{} and is 80 in \quantgan{}. We calculate $M=512$ simulated paths of length $T=2560$ for evaluation. In the correlation scores, we let $\delta=64$. The loss for training \quantgan{} is the loss of the original \gan{} in Equation \eqref{eq:loss_gan}. We use the loss functions of the \wgangp{} in Equation \eqref{eq:loss_wgan_gp} to train \tagan{} and \ttgan{}.  

The results of \tagan{} with and without PCA, \ttgan{} with and without the return feature, and \quantgan{}, are summarized in Table \ref{tab:option_result}. The good candidates, which are \tagan{} with PCA, \ttgan{} with the return feature, and \quantgan{} are further illustrated in Figure \ref{fig:tagan_option}, \ref{fig:ttgan_option} and \ref{fig:quantgan_option}. Here are some key points of results:
\begin{itemize}
	\item Only \ttgan{} is improved by the additional return feature. It is not a surprise to see \ttgan{} can accept the additional return feature, since it accepts both log-prices and log returns for the S\&P 500 index simulation. The additional return feature improves the score of autocorrelation and cross-correlation, and facilitates the \gan{} to learn the no-arbitrage condition.
	\item Only \tagan{} is improved by PCA. If a \gan{} is able to generate option surfaces by means of principle components, that will significantly reduce the score of cross-correlation and reduce the rate that the output needs to be modified by the no-arbitrage condition.
	\item In Figure \ref{fig:option_diffacf}, we show examples of autocorrelation of log-volatility returns from the three \gans{}. There are huge fluctuations in the autocorrelation of \quantgan{}. Also, some fluctuations are observed at the small time lags in the autocorrelation of \tagan{}. In contrast, the autocorrelation of \ttgan{} is flat. It means the attention layer is better at generating sequences with smooth autocorrelation, which matches the results of the S\&P 500 index simulation.
\end{itemize}
   
\begin{table}[h]
\centering
\begin{small}
\begin{tabular}{cccccc}\hline
	scores & \begin{tabular}{c}\tagan{}\\(w/o PCA)\end{tabular} & \begin{tabular}{c}\tagan{}\\(w/ PCA)\end{tabular} & \begin{tabular}{c}\ttgan{}\\(w/o returns)\end{tabular} & \begin{tabular}{c}\ttgan{}\\(w/ returns)\end{tabular} & \quantgan{} \\\hline
	$W_1^{(1)}$ & 1.788e-02 & 1.651e-02 & 1.239e-02 & 1.512e-02 & 1.355e-02\\
	skewness & 2.434e-01 & 2.450e-01 & 2.077e-01 & 8.204e-02 & 8.560e-02\\
	kurtosis & 6.052e-01 & 2.710e-01 & 5.212e-01 & 5.607e-01 & 4.065e-01\\
	ACF & 3.065e-01 & 3.444e-01 & 4.359e-01 & 1.845e-01 & 1.754e-01\\
	ACF$^{(\text{r})}$ & 3.601e-01 & 2.580e-01 & 3.727e-01 & 2.667e-01 & 8.683e-01\\
	cross-corr & 4.883e-01 & 1.016e-01 & 6.027e-01 & 2.618e-01 & 2.284e-01\\
	arbitrage rate & 30.10\% & 1.55\% & 21.16\% & 8.88\% & 12.86\% \\
\hline
\end{tabular}
\end{small}
\caption{Scores of the S\&P 500 index option surface simulation.}
\label{tab:option_result}
\end{table}

\section{Conclusion}\label{sec:conclusion}
In this paper, we first define the generative model of time series, distinguish it from the generator of fixed dimension distributions. We then propose two \gans{}, the temporal attention \gan{} and the temporal transformer \gan{}, based on the causal attention layer, which is able to increase the receptive field size without introducing more parameters. We have successfully trained the temporal transformer \gan{} using around 3000 samples of financial time series with the help of sparse attention, despite the fact that both \gans{} and transformers are notoriously known for being difficult to train.  

In the numerical experiments, we compare the two proposed \gans{} with \quantgan{} for the stock index and option surface simulation and evaluate the results with the scores based on the stylized facts. The proposed \gans{} are able to replicate the distribution, the heavy tails and the long-range dependencies, as well as the cross-correlation in the multivariate case. Specifically, the attention-based \gans{} show the following advantages:
\begin{itemize}
	\item The \ttgan{} tends to generate smoother autocorrelation of returns. However, the convolution-based \quantgan{} tends to overfit autocorrelation curves. For option surface simulation, the \quantgan{} even fails to replicate the autocorrelation of volatility returns. The \tagan{}, as a mixture of convolutions and attention, lies between \ttgan{} and \quantgan{}.
	\item The transformer discriminator in \ttgan{} is more flexible such that it can accept both level and return features. We can make use of its ability to process features of different scales to improve the performance of \gans{}. 
	\item The \tagan{} is able to learn and generate samples in the space of principal components, which makes it possible to simulate time series in higher-dimensional spaces utilizing PCA.
	\item The receptive field size of the attention-based \gans{} is not bounded by the number of parameters or the network depth. This is useful especially when the size of real data is limited and a large number of parameters would lead to overfitting.
\end{itemize}

The generative models discussed in the paper are all unconditional models, which generate time series given noise series. In the future, it would be interesting to compare the performance of unconditional models and conditional models, which generate future time series given noise as well as historical time series. The conditional models are trickier for the following reasons:
\begin{itemize}
	\item The conditional models need to learn the conditional distribution given history time series, which is generally more complex than the unconditional distribution.
	\item The input of the unconditional model is random noise generated during training. Thus, the unconditional model would not memorize the input. However, the conditional model can easily remember real data and perform extremely well when real data is used as the condition input. When the conditional model uses the time series generated by itself as the condition input and tries to prolong the generated time series, its performance could be much worse. For that reason, we need additional techniques to deal with overfitting.
\end{itemize}      

\clearpage  
\bibliographystyle{abbrv}
\bibliography{references.bib}

\clearpage
\appendix
\section{Numerical experiment results}\label{app:results}
\begin{figure}[h]
    \centering
 	\includegraphics[width = 0.9\textwidth]{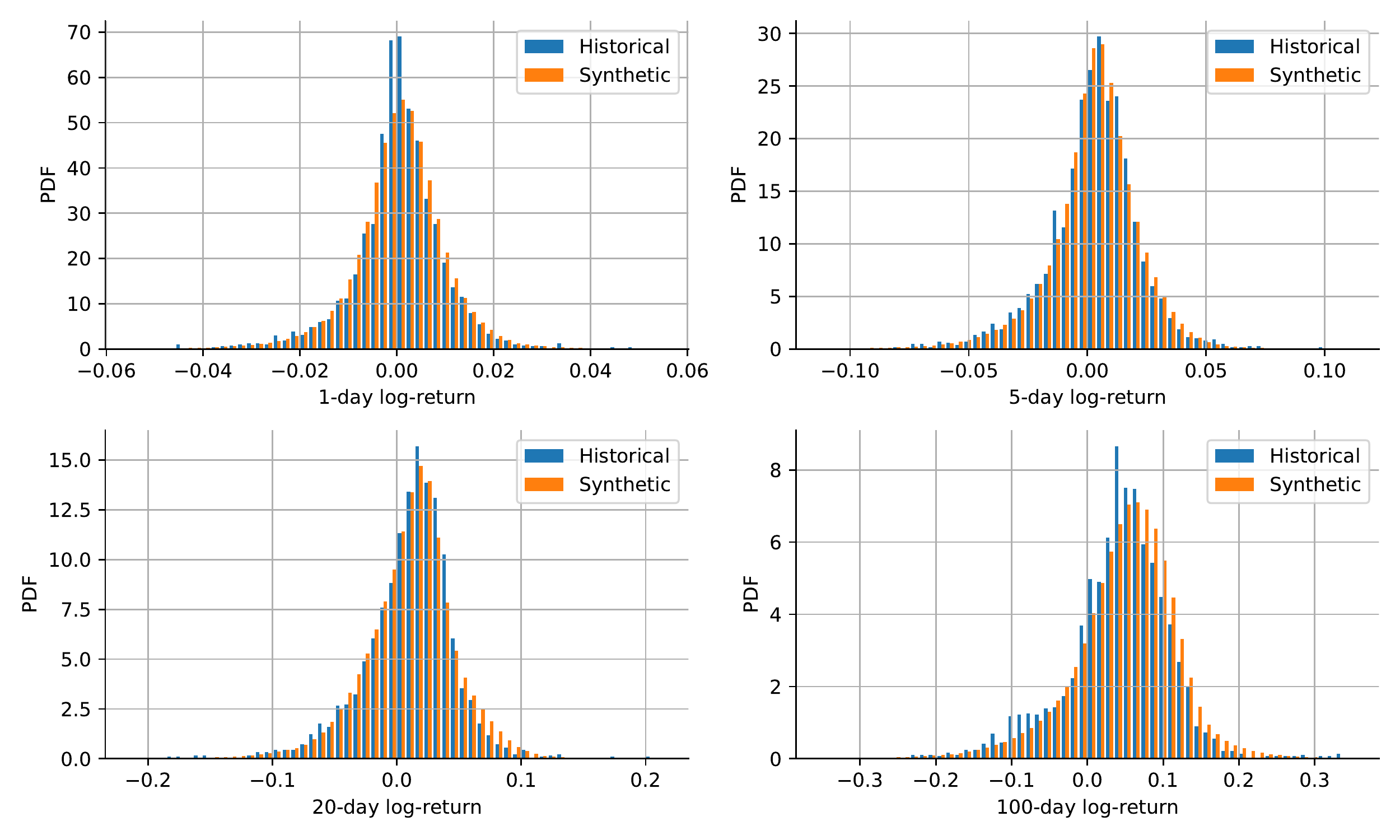}\\
 	(a)\\
 	\includegraphics[width = 0.9\textwidth]{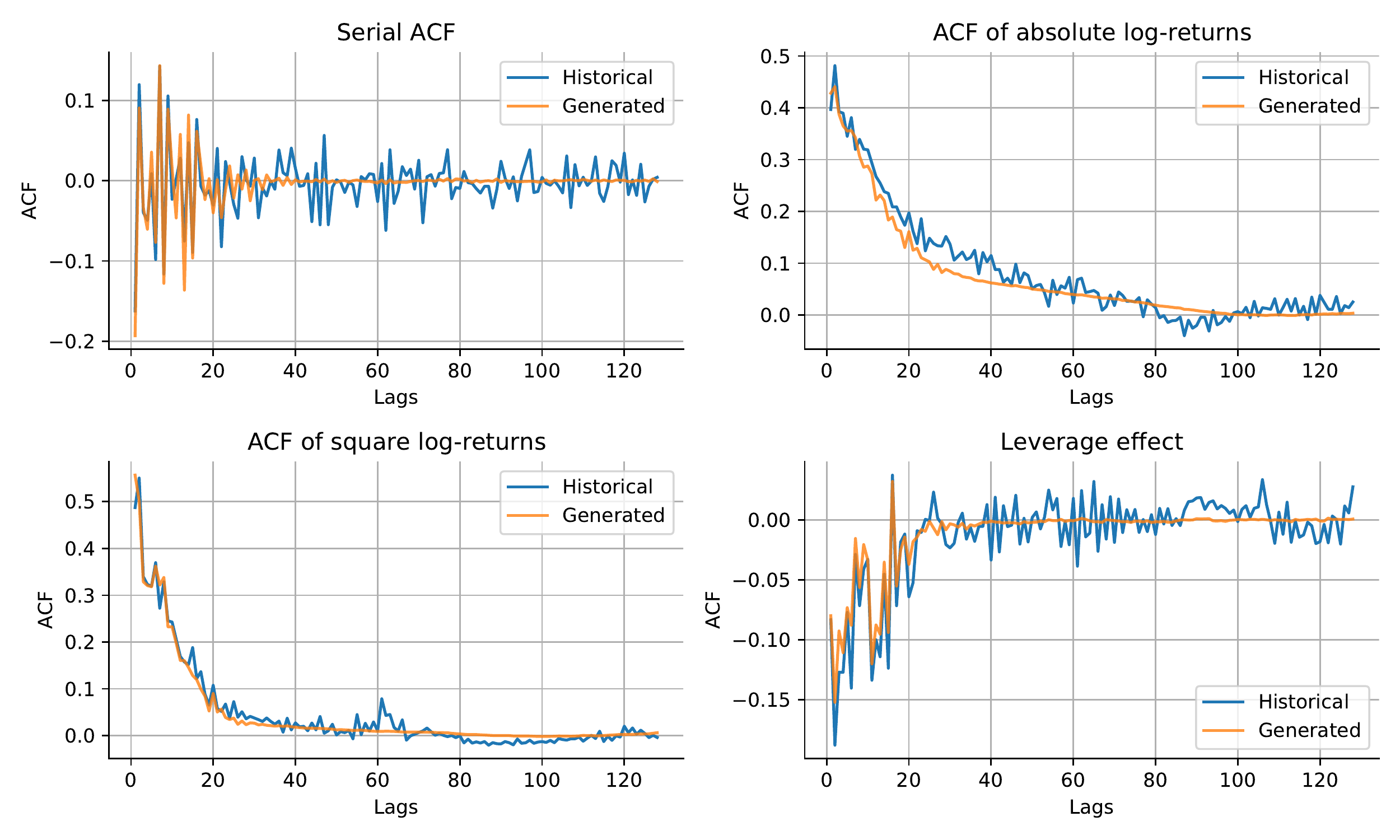}\\
 	(b)
 	\caption{\tagan{} simulation results of the S\&P 500 index. (a) Comparison of the generated and historical densities of log returns. (b) Comparison of the generated mean autocorrelation and historical autocorrelation of daily log returns.}
    \label{fig:tagan_index}
\end{figure}

\begin{figure}[h]
    \centering
 	\includegraphics[width = 0.9\textwidth]{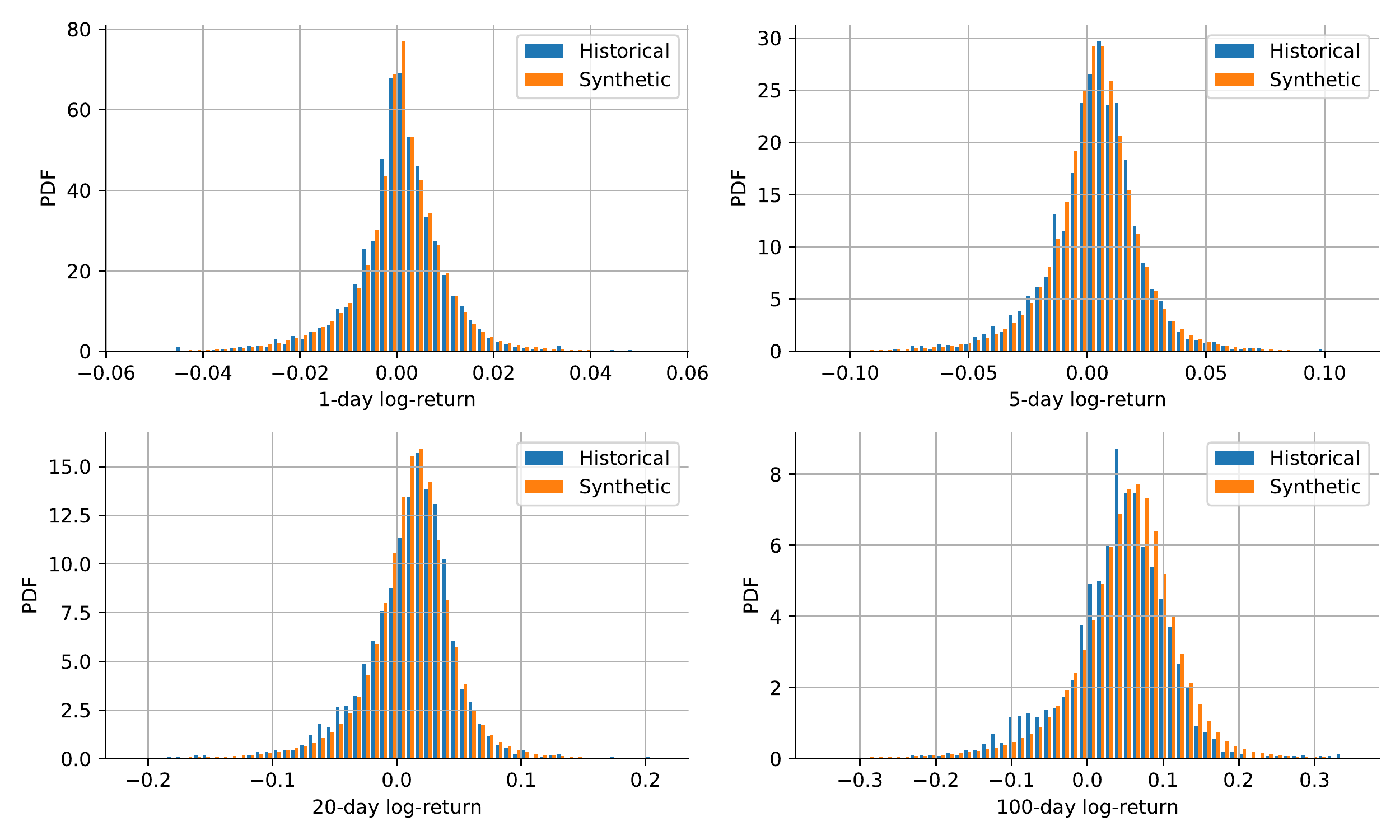}\\
 	(a)\\
 	\includegraphics[width = 0.9\textwidth]{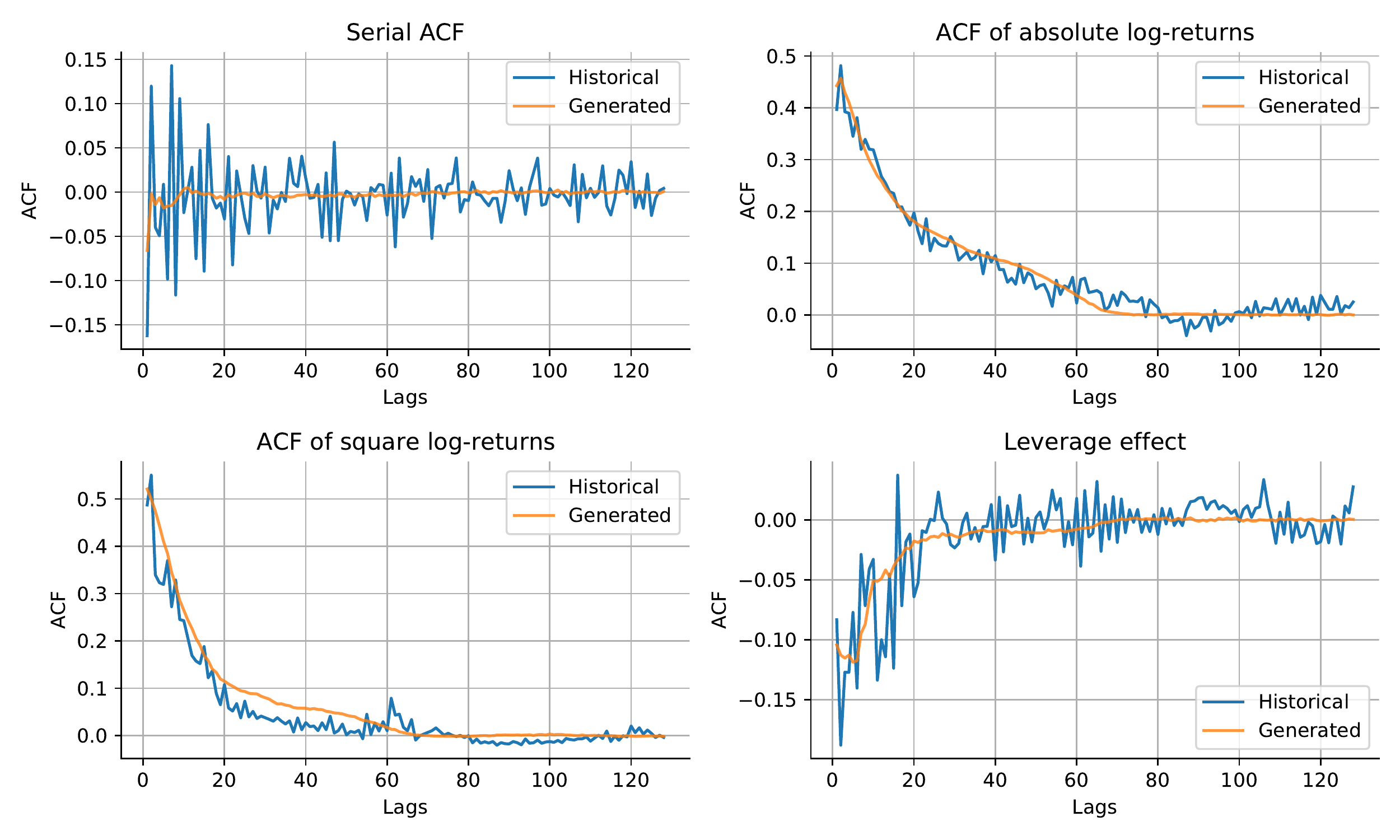}\\
 	(b)
 	\caption{\ttgan{} simulation results of the S\&P 500 index. (a) Comparison of the generated and historical densities of log returns. (b) Comparison of the generated mean autocorrelation and historical autocorrelation of daily log returns.}
    \label{fig:ttgan_index}
\end{figure}

\begin{figure}[h]
    \centering
 	\includegraphics[width = 0.9\textwidth]{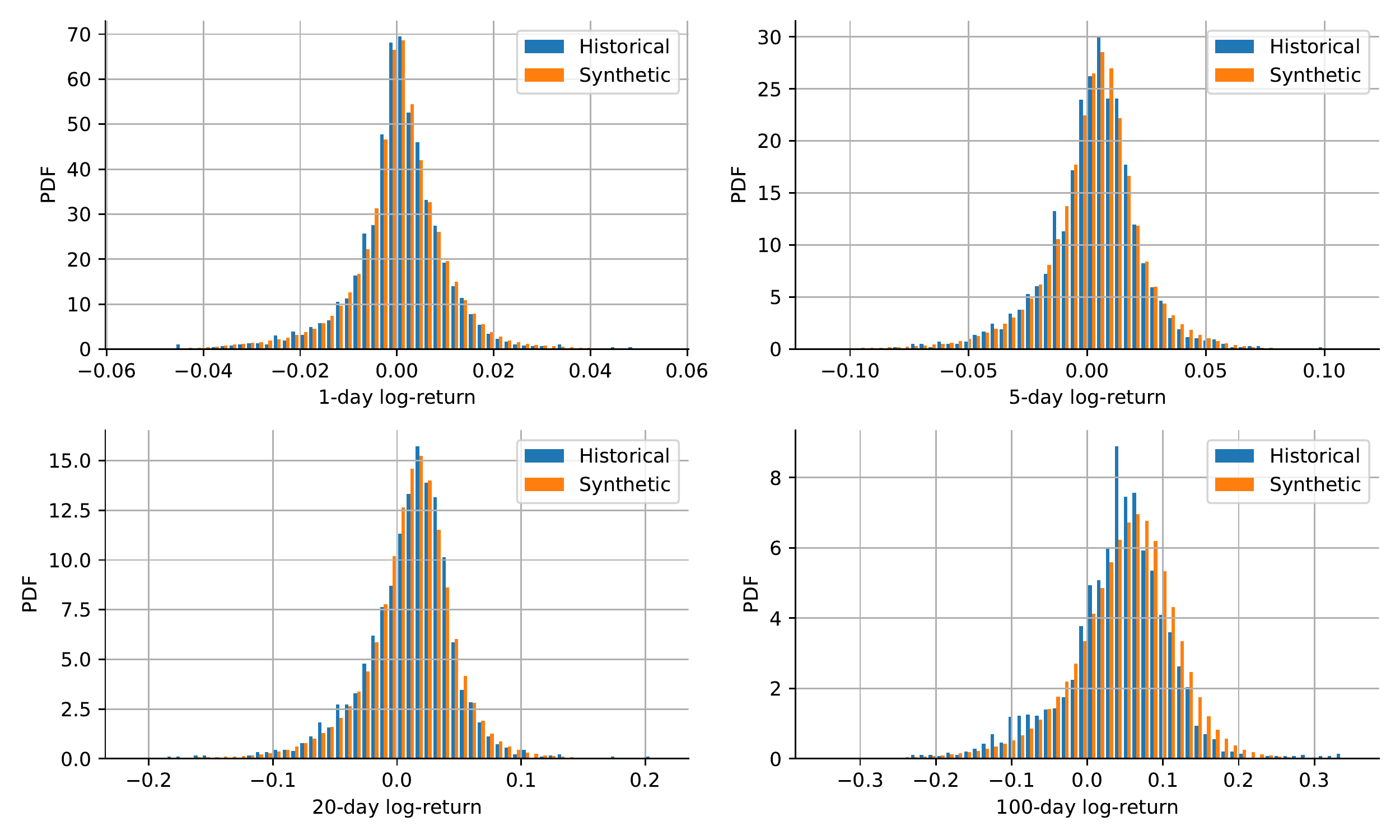}\\
 	(a)\\
 	\includegraphics[width = 0.9\textwidth]{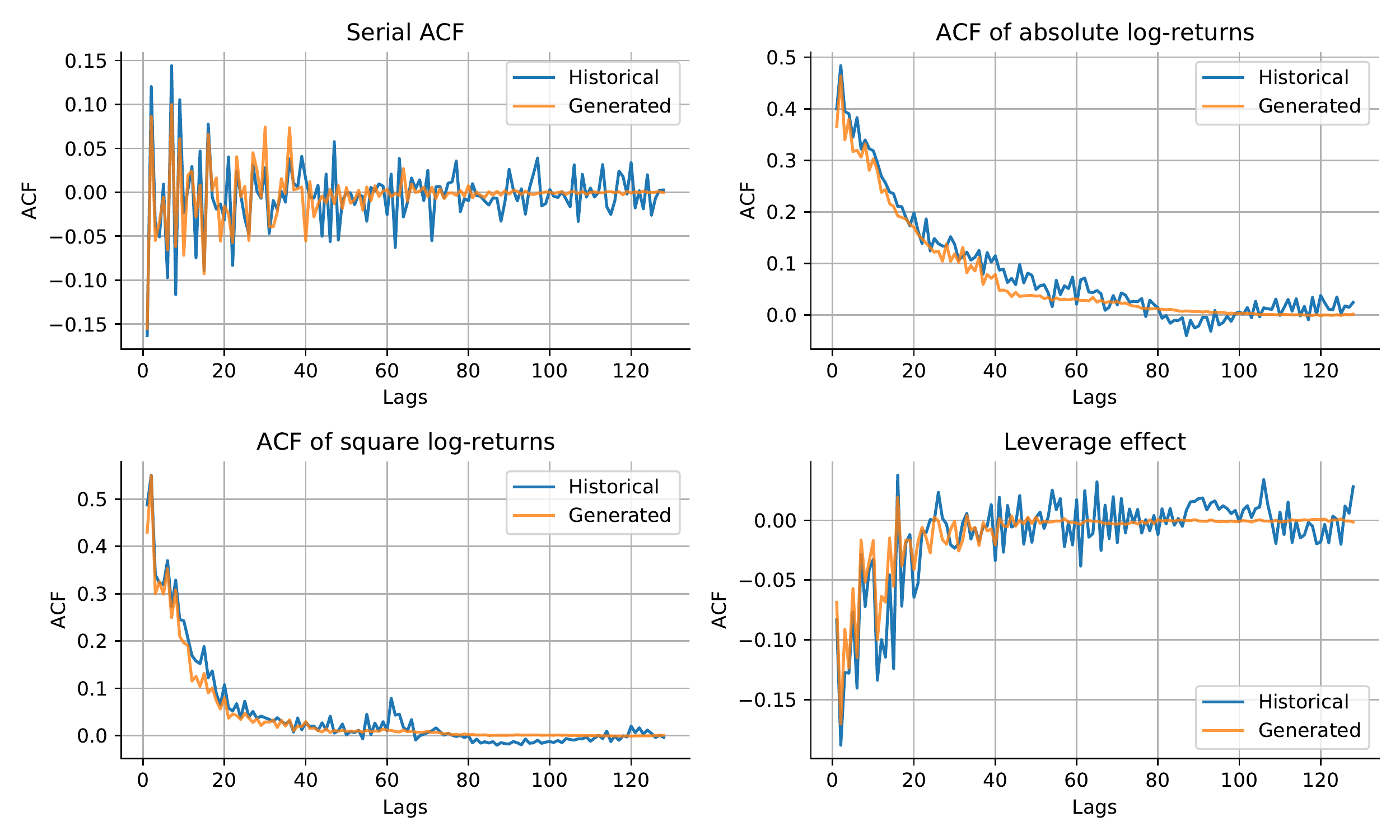}\\
 	(b)
 	\caption{\quantgan{} simulation results of the S\&P 500 index. (a) Comparison of the generated and historical densities of log returns. (b) Comparison of the generated mean autocorrelation and historical autocorrelation of daily log returns.}
    \label{fig:quantgan_index}
\end{figure}

\begin{figure}[h]
    \centering
 	\includegraphics[width = 0.8\textwidth]{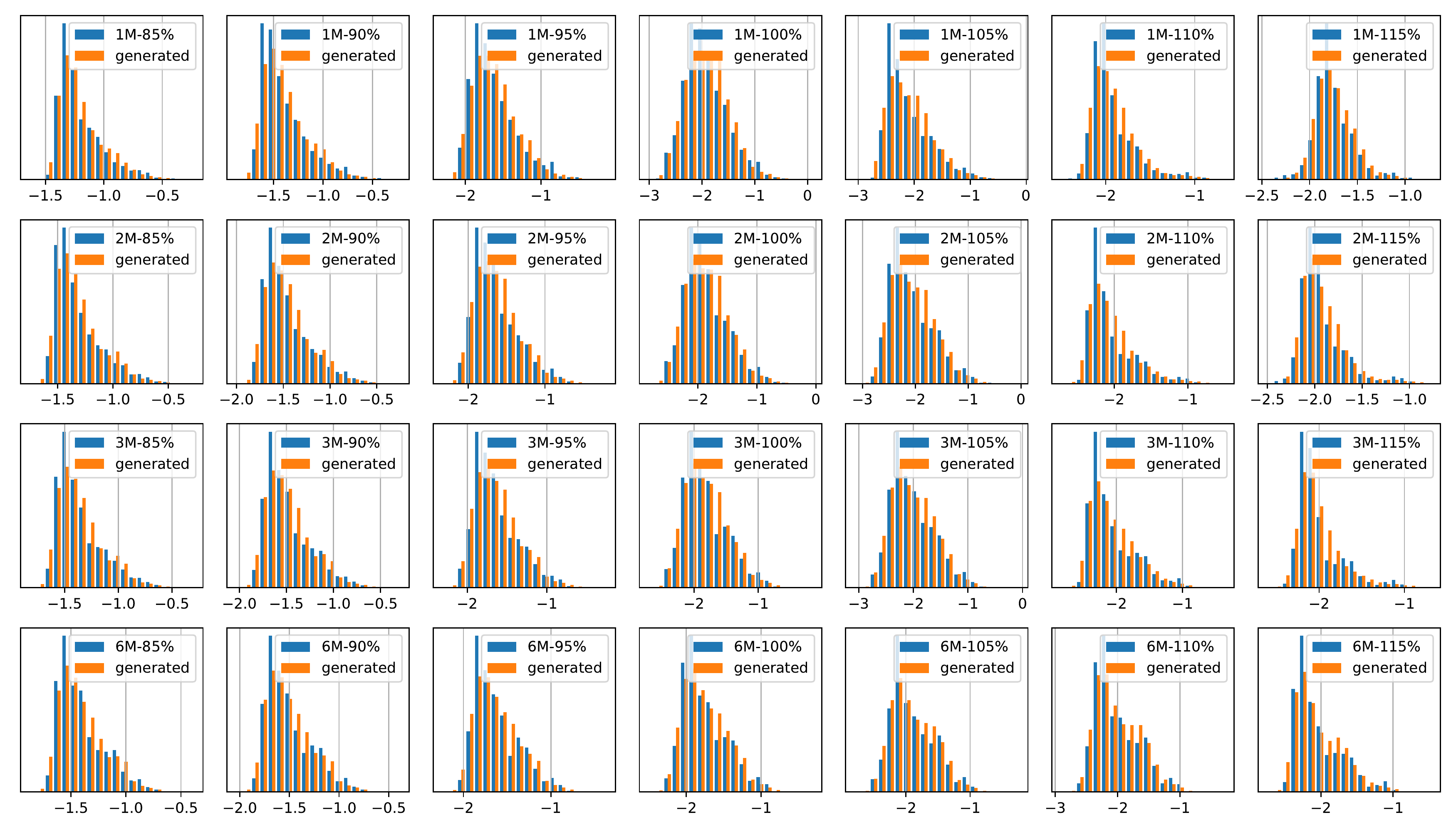}\\
 	(a)\\
 	\includegraphics[width = 0.8\textwidth]{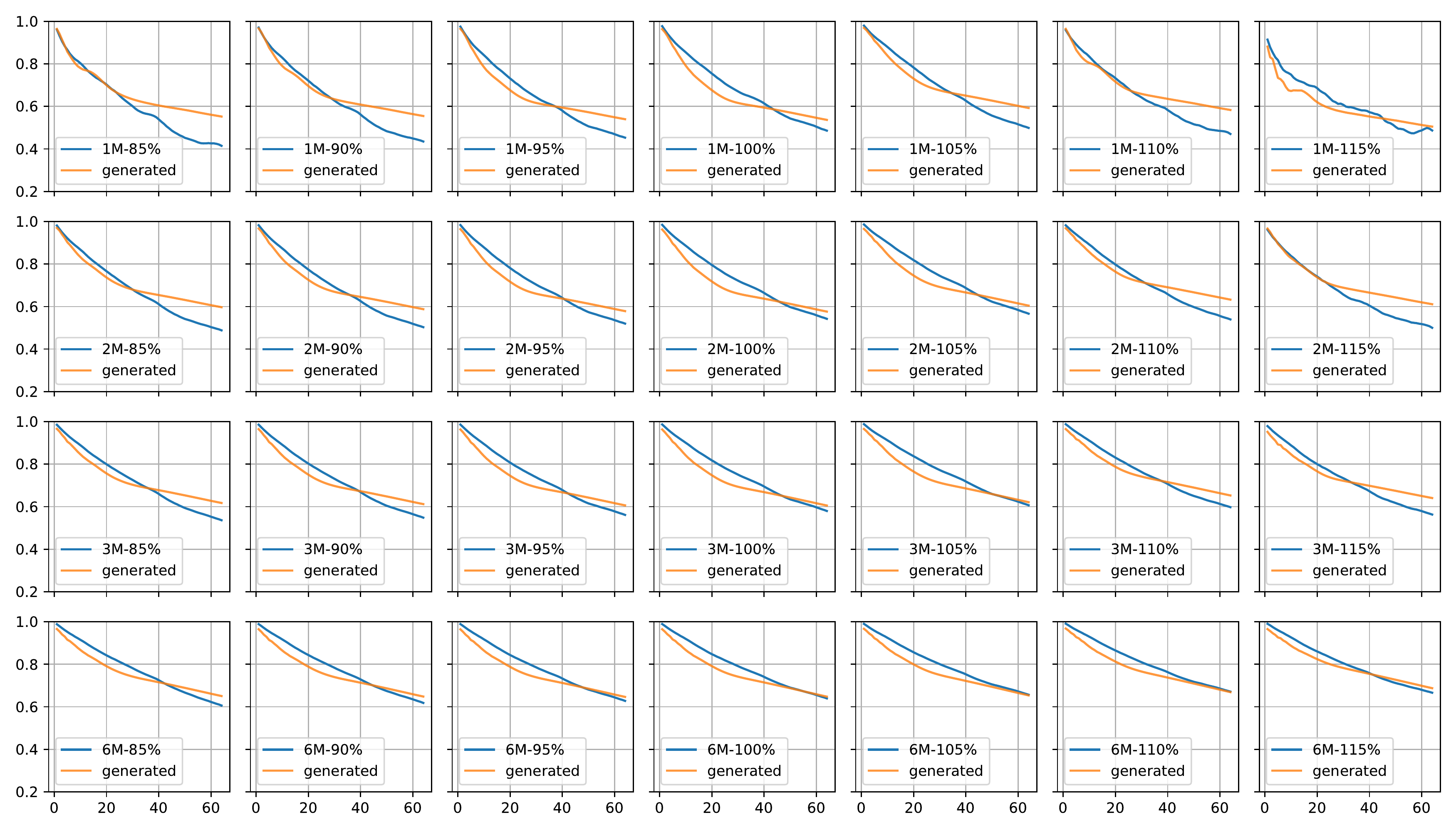}\\
 	(b)\\
 	\includegraphics[width = 0.5\textwidth]{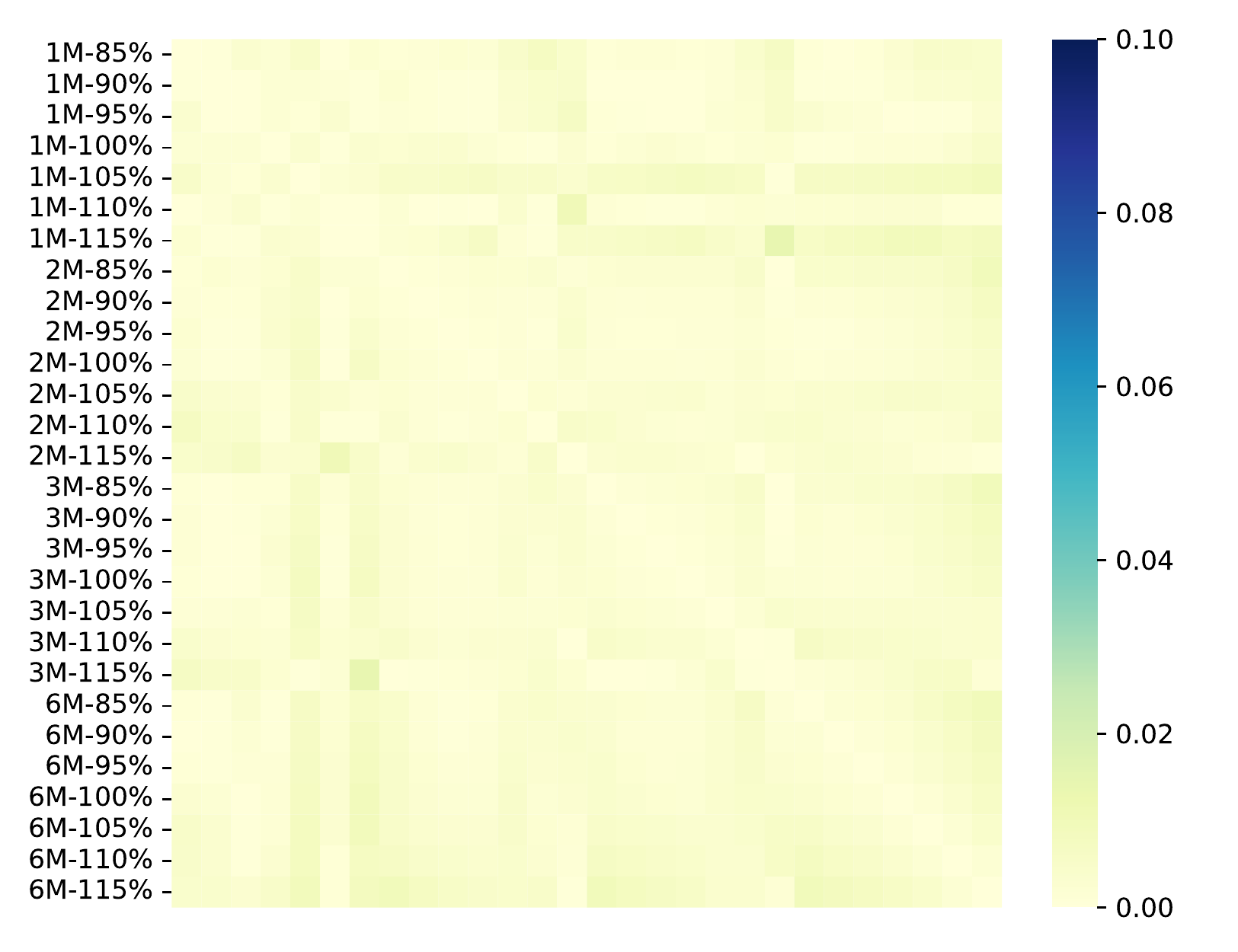}\\
 	(c)
 	\caption{\tagan{} simulation results of the S\&P 500 index options. (a) Comparison of the generated and historical densities of log-volatilities. (b) Comparison of the generated mean autocorrelation and historical autocorrelation of log-volatilities. (c) Difference of the generated and historical cross-correlation of log-volatilities $\vert\Sigma_{\vx}-\Sigma_{\vy}\vert$.}
    \label{fig:tagan_option}
\end{figure}

\begin{figure}[h]
    \centering
 	\includegraphics[width = 0.8\textwidth]{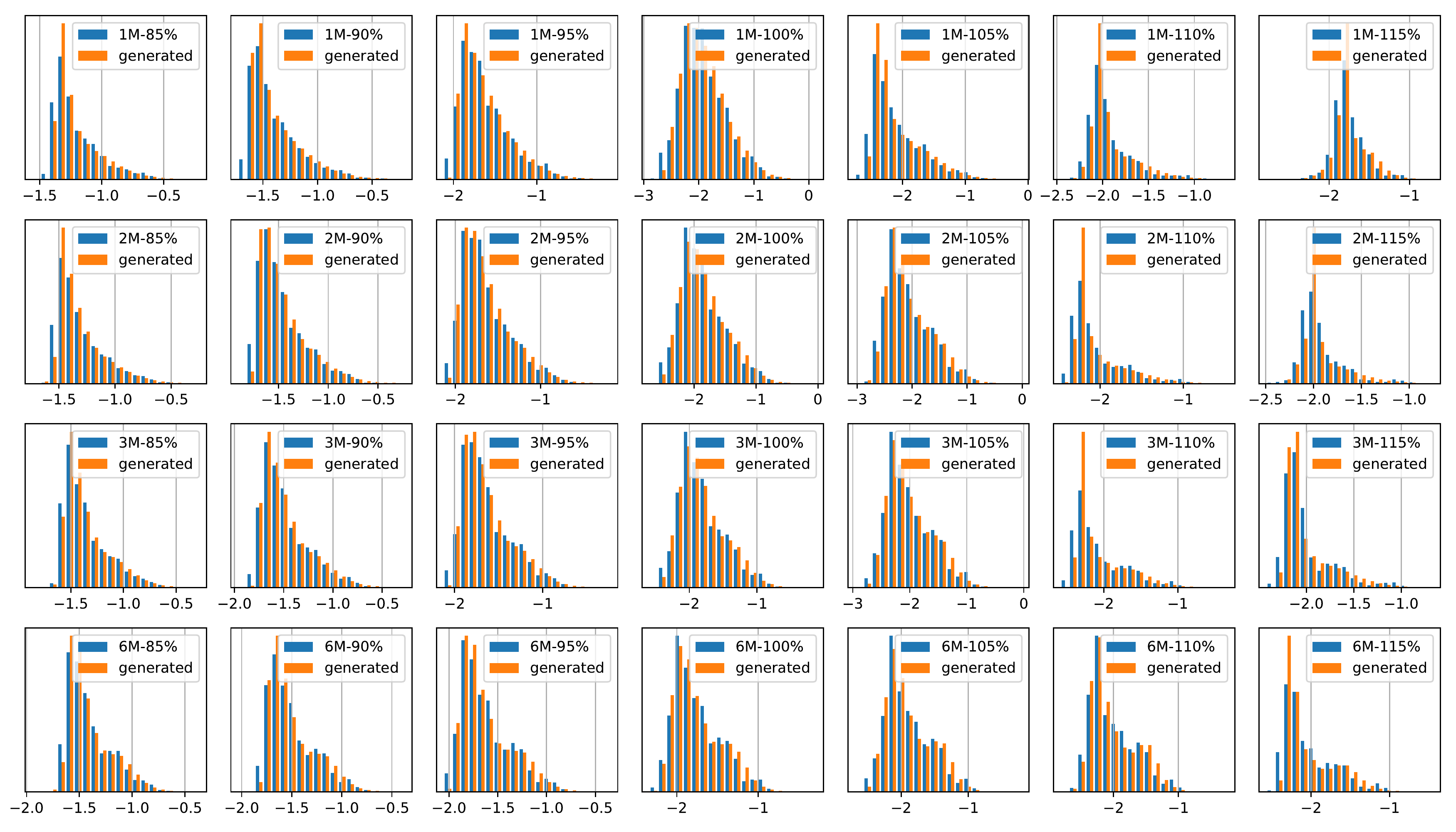}\\
 	(a)\\
 	\includegraphics[width = 0.8\textwidth]{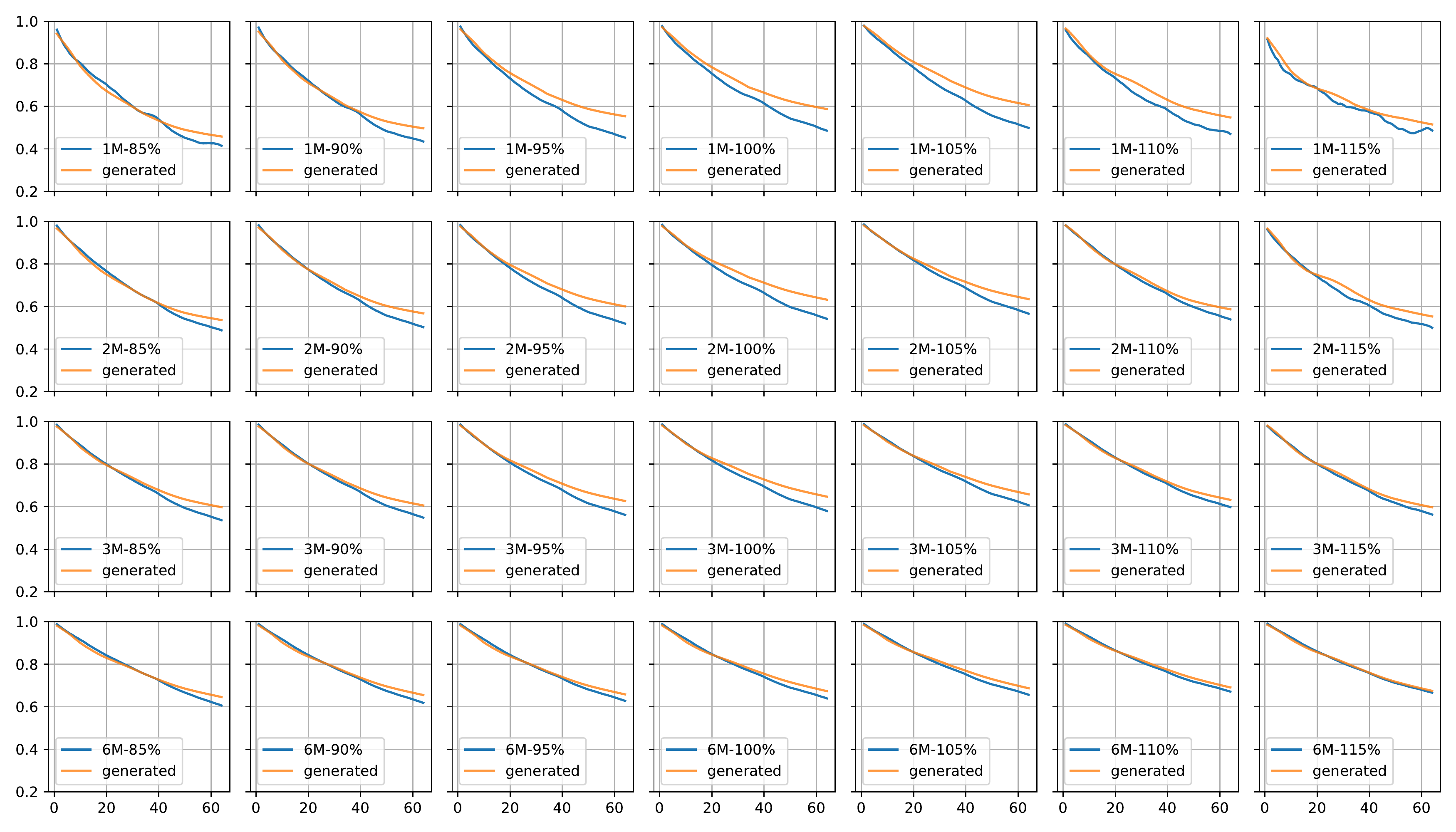}\\
 	(b)\\
 	\includegraphics[width = 0.5\textwidth]{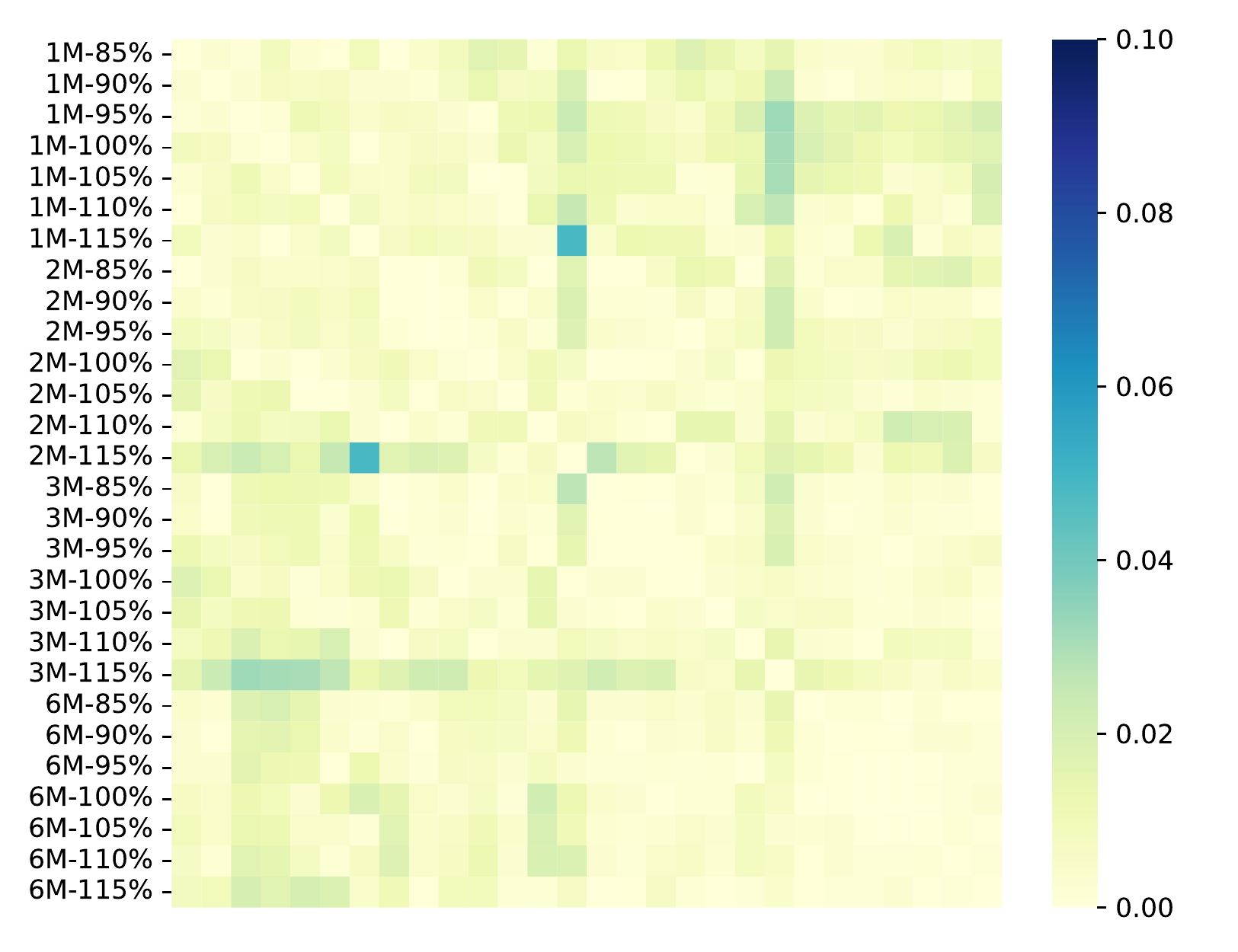}\\
 	(c)
 	\caption{\ttgan{} simulation results of the S\&P 500 index options. (a) Comparison of the generated and historical densities of log-volatilities. (b) Comparison of the generated mean autocorrelation and historical autocorrelation of log-volatilities. (c) Difference of the generated and historical cross-correlation of log-volatilities $\vert\Sigma_{\vx}-\Sigma_{\vy}\vert$.}
    \label{fig:ttgan_option}
\end{figure}

\begin{figure}[h]
    \centering
 	\includegraphics[width = 0.8\textwidth]{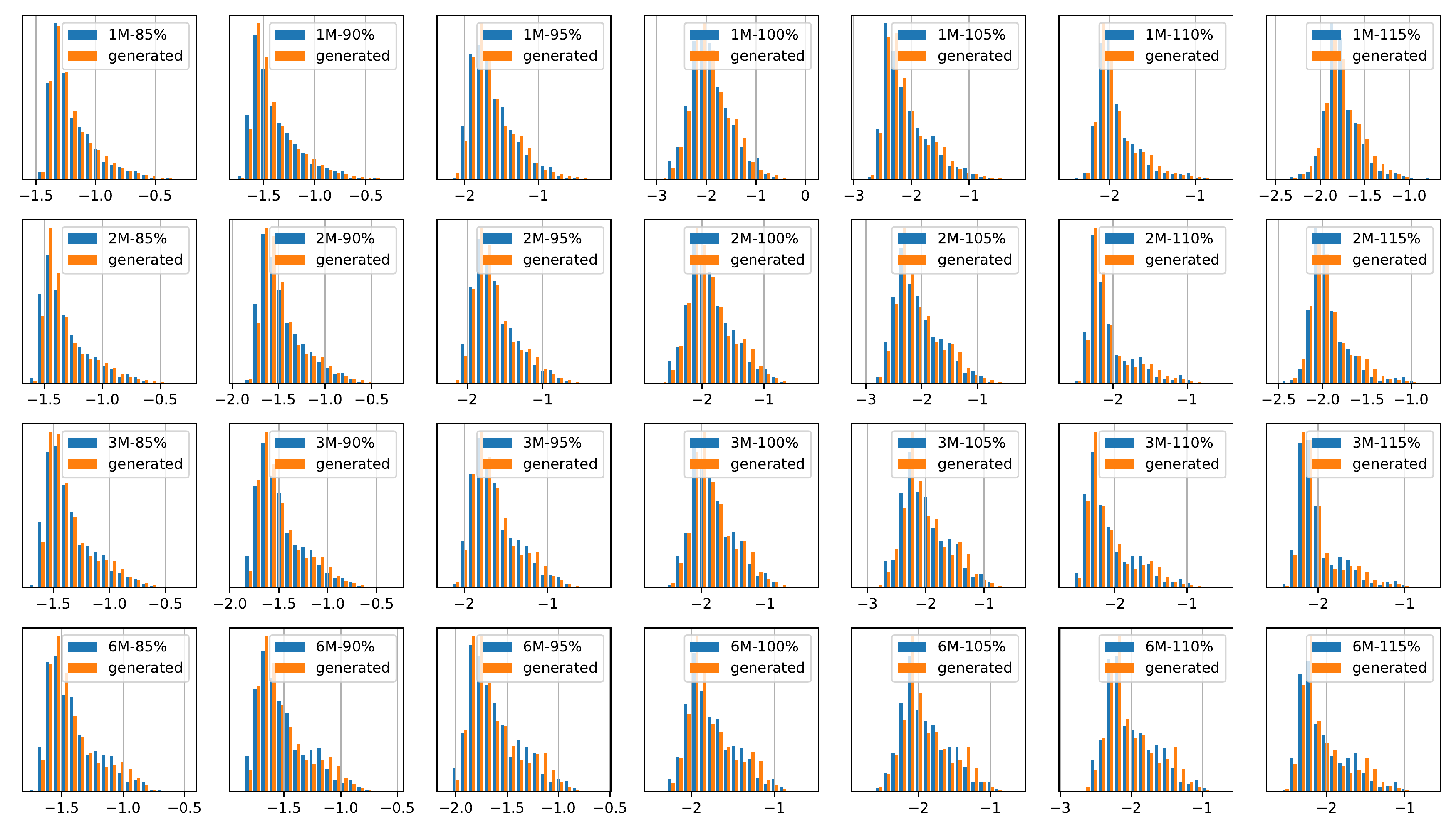}\\
 	(a)\\
 	\includegraphics[width = 0.8\textwidth]{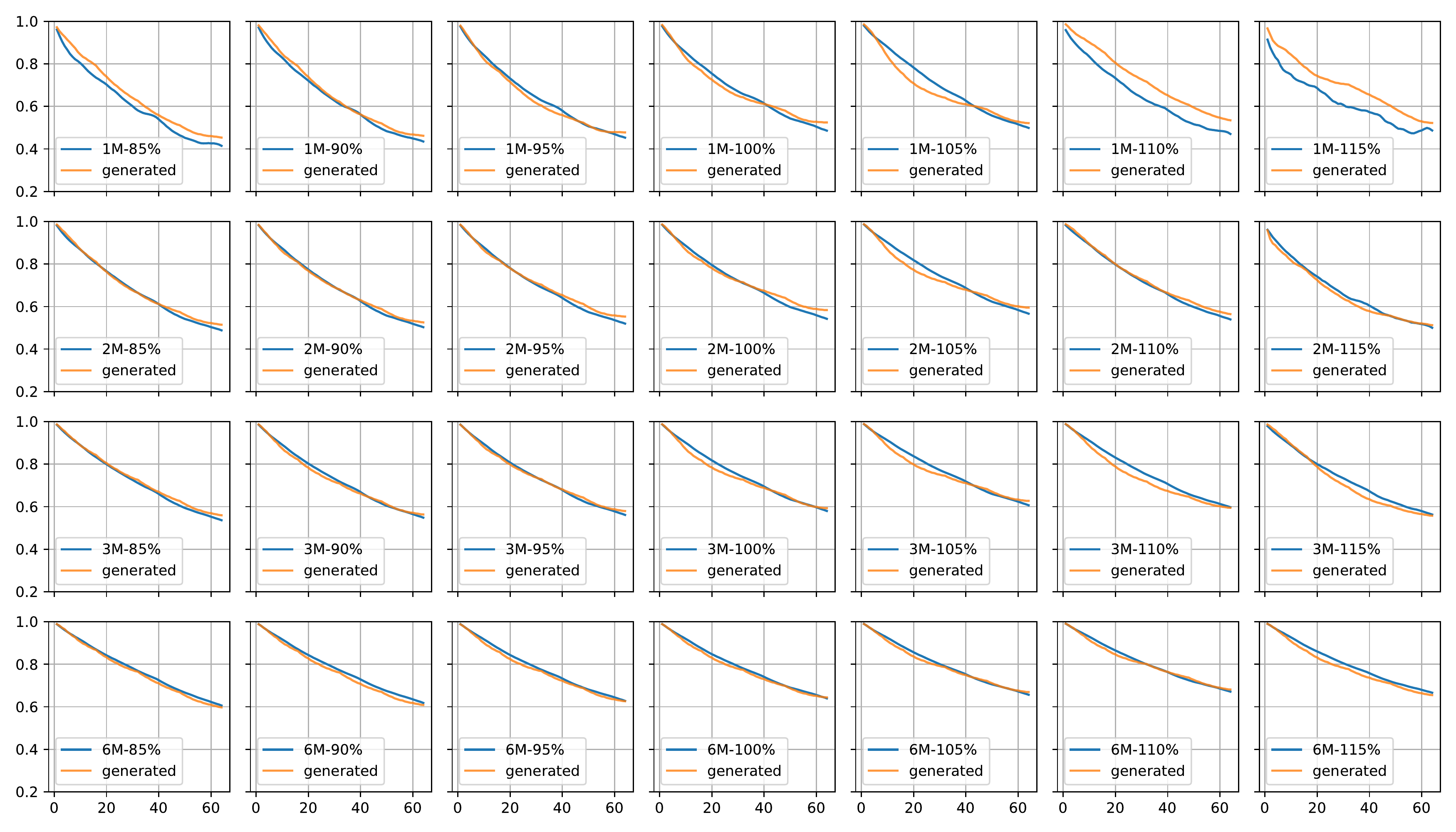}\\
 	(b)\\
 	\includegraphics[width = 0.5\textwidth]{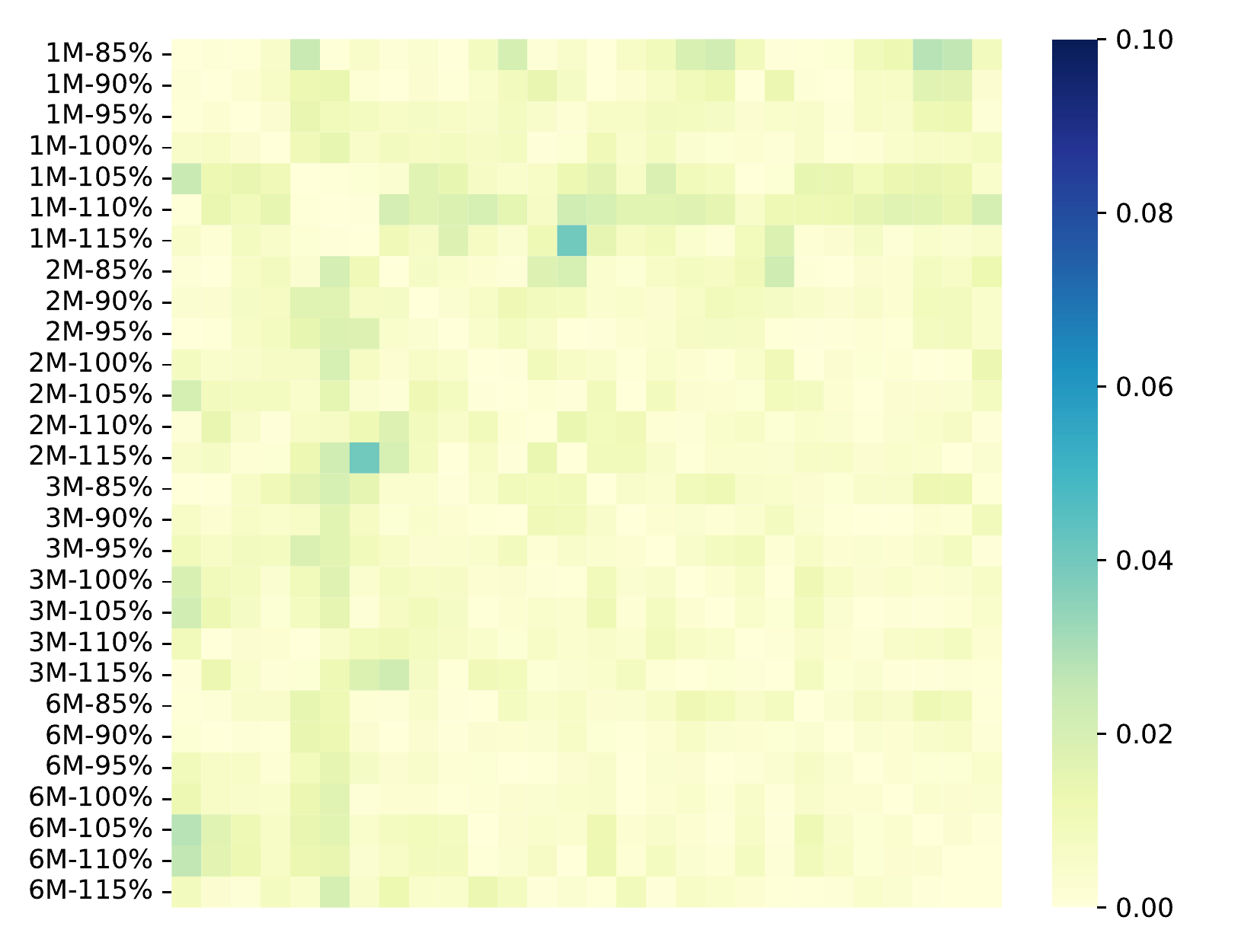}\\
 	(c)
 	\caption{\quantgan{} simulation results of the S\&P 500 index options. (a) Comparison of the generated and historical densities of log-volatilities. (b) Comparison of the generated mean autocorrelation and historical autocorrelation of log-volatilities. (c) Difference of the generated and historical cross-correlation of log-volatilities $\vert\Sigma_{\vx}-\Sigma_{\vy}\vert$.}
    \label{fig:quantgan_option}
\end{figure}

\clearpage

\begin{figure}[h]
    \centering
    \begin{tabular}{ccc}
    	\includegraphics[width = 0.3\textwidth]{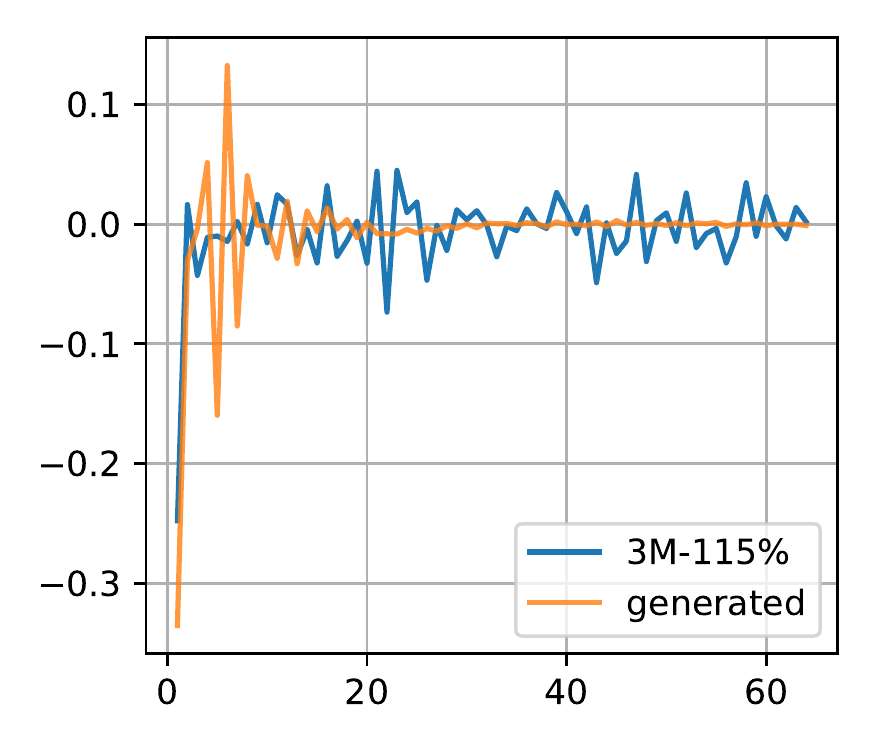} & \includegraphics[width = 0.3\textwidth]{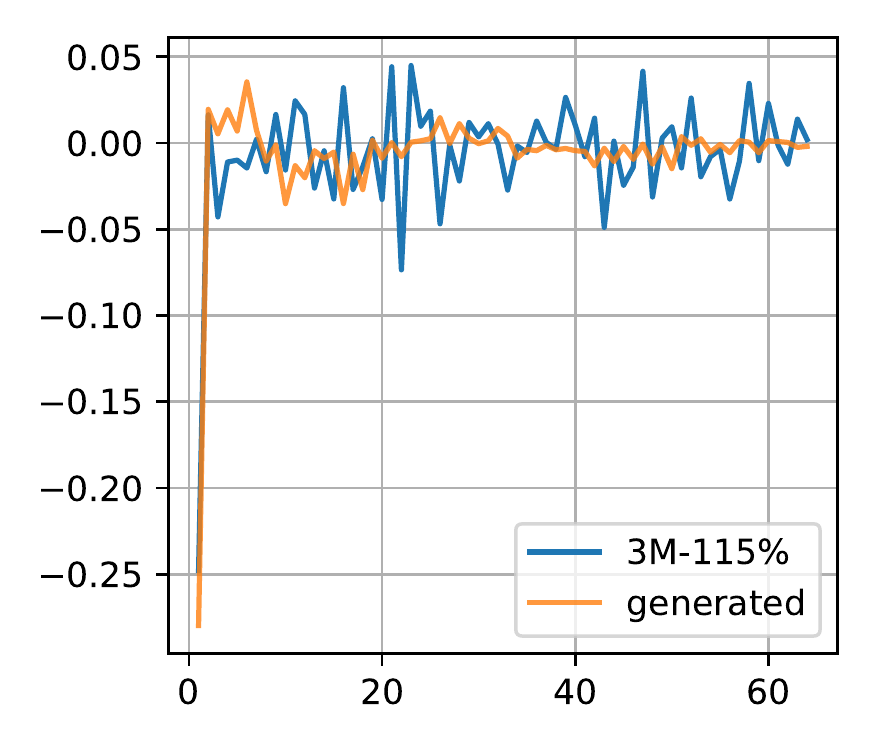} & \includegraphics[width = 0.3\textwidth]{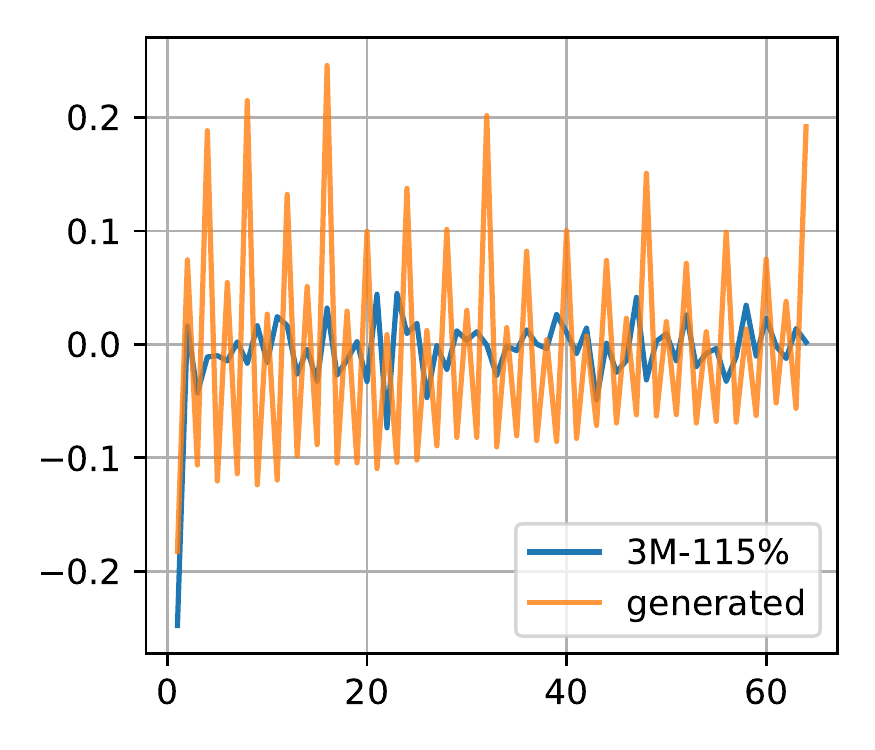} \\
 	(a) & (b) & (c)\\
    \end{tabular}
 	\caption{Example of the generated mean autocorrelation and historical autocorrelation of log-volatility returns of the S\&P 500 index options. (a) \tagan{}. (b) \ttgan{}. (c) \quantgan{}.}
    \label{fig:option_diffacf}
\end{figure}

\section{Arbitrage-free option surface}\label{app:no_arbitrage}
In \cite{buehler2017discrete}, the authors give the condition that a call option surface is arbitrage-free. Suppose we have the set of relative strikes
$$\mathcal{K}=\{K_0,K_1,K_2,\dots,K_{N_K},K_{N_K+1}\}$$
where $K_i<K_{i+1},0\leq i\leq N_K$ and the set of maturities
$$\mathcal{M}=\{M_1,M_2,\dots,M_{N_M}\}$$ where $M_j<M_{j+1},1\leq j\leq N_M-1$. Let $C_{i,j}$ be the call price at strike $K_i$ and maturity $M_j$. $K_0$ is sufficiently small and $K_{N_K+1}$ is sufficiently large, so that we have $C_{0,j}=1-K_0$ and $C_{N_K+1,j}=0$. Then the variables $\{C_{i,j},1\leq i\leq N_K,1\leq j\leq N_M\}$ need to satisfy the following conditions:
\begin{align}
\begin{cases}
C_{1,j}\geq 1-K_1, & \forall 1\leq j\leq N_M\\
C_{N_K,j}\geq 0, & \forall 1\leq j\leq N_M \\
C_{i,j}\geq C_{i,j-1}, &\forall 1\leq i\leq N_K,2\leq j\leq N_M \\
\frac{C_{i,j}-C_{i-1,j}}{K_i-K_{i-1}}\leq \frac{C_{i+1,j}-C_{i,j}}{K_{i+1}-K_{i}}, &\forall 1\leq i\leq N_K, 1\leq j\leq N_M
\end{cases}\label{eq:no-arbitrage}
\end{align}
Suppose $\{\hat{C}_{i,j},1\leq i\leq N_K,1\leq j\leq N_M\}$ is the call option surface that does not satisfy the conditions, we use the following linear programming to solve the closest arbitrage-free surface.
\begin{align*}
	\min_{C_{i,j},1\leq i\leq N_K,1\leq j\leq N_M}&\sum_{i=1}^{N_K}\sum_{j=1}^{N_M}\vert \hat{C}_{i,j}-{C}_{i,j}\vert\\
	\text{such that}\,&\,\text{the constraints in}\,\eqref{eq:no-arbitrage}\,\text{are satisfied.}
\end{align*}
If we start from the volatility surface, we calculate the call options and detect any violations of the constraints in \eqref{eq:no-arbitrage}. If so, we perform the linear programming to remove the arbitrages and calculate the corrected implied volatilities.

\section{Losses of \gans{}}\label{app:gan_loss}
In \cite{goodfellow2014generative}, the authors proposed the original loss of \gans{}:
\begin{align*}
	\min_{\theta_G}\max_{\theta_D} \E_{\mX} \ln(D(\mX;\theta_D))+\E_{\mZ}\ln(1-D(G(\mZ;\theta_G);\theta_D))
\end{align*}
where the discriminator is $D(\cdot;\theta_D):\R^{l\times d}\rightarrow (0,1)$ and the output of the discriminator stands for the probability that its input is considered real data. Thus the losses for the generator and the discriminator are 
\begin{align*}
	&\min_{\theta_G}\E_{\mZ}\ln(1-D(G(\mZ;\theta_G);\theta_D))\\
	&\min_{\theta_D} -\E_{\mX}\ln(D(\mX;\theta_D))-\E_{\mZ}\ln(1-D(G(\mZ;\theta_G);\theta_D))
\end{align*}
respectively. In practice, $D(G(\mZ;\theta_G);\theta_D)$ is close to 0 in the beginning since the generator has not learned anything. The gradient of $\ln(1-D(G(\mZ;\theta_G);\theta_D))$ is small and convergency would be slow. So the loss for the generator is replaced with 
\begin{align*}
	\min_{\theta_G}-\E_{\mZ}\ln(D(G(\mZ;\theta_G);\theta_D)).
\end{align*}
In this paper, we use the discriminator $D(\cdot;\theta_D):\R^{l\times d}\rightarrow \R$ to include the case of \wgangp{}. So the sigmoid function $\sigma(D(\cdot;\theta_D))$ stands for the probability. Thus the original losses of \gans{} are written as 
\begin{align*}
	&\min_{\theta_G}\E_{\mZ}-\ln(\sigma(D(G(\mZ;\theta_G);\theta_D)))\\
	&\min_{\theta_D} \E_{\mX,\mZ}-\ln(\sigma(D(\mX;\theta_D)))-\ln(1-\sigma(D(G(\mZ;\theta_G);\theta_D))).
\end{align*}

The Wasserstein \gan{} in \cite{arjovsky_wasserstein_2017} approaches the loss of \gans{} in a different way. It tries to minimize the Wasserstein-1 distance between the real distribution and the generated distribution:
\begin{align*}
	W(\pr_r,\pr_g)=\inf_{\gamma\in\Pi(\pr_r,\pr_g)}\E_{(\mX,\mY)\sim\gamma}\Vert \mX-\mY\Vert
\end{align*}
where $\pr_r$ is the real distribution of $\mX$ and $\pr_g$ is the generated distribution of $G(\mZ;\theta_G)$, and $\Pi(\pr_r,\pr_g)$ denotes the set of all joint distributions $\gamma$ whose marginals are respectively $\pr_r$ and $\pr_g$. $\Vert \cdot \Vert$ is the Frobenius norm. Then they make use of the Kantorovich-Rubinstein duality \cite{villani2009optimal} to get 
\begin{align*}
	W(\pr_r,\pr_g)=\sup_{\Vert f\Vert_L\leq 1}\E_{\mX}f(\mX)-\E_{\mZ}f(G(\mZ;\theta_G))
\end{align*}
where $\Vert f\Vert_L\leq 1$ means the 1-Lipschitz function $f$. For a differentiable $f$, this means $\Vert \nabla f\Vert\leq 1$. The discriminator $D(\cdot;\theta_D):\R^{l\times d}\rightarrow \R$ is used to approximate the function $f$ that reaches the supremum and its loss is 
\begin{align*}
	\min_{\theta_D} -\E_{\mX}D(\mX;\theta_D)+\E_{\mZ}D(G(\mZ;\theta_G);\theta_D).
\end{align*}
The generator tries to minimize the Wasserstein-1 distance, which means
\begin{align*}
	\min_{\theta_G}\E_{\mZ}-D(G(\mZ;\theta_G);\theta_D).
\end{align*}
Note that the discriminator needs to be 1-Lipschitz continuous such that the Kantorovich-Rubinstein duality holds. Thus the authors in \cite{gulrajani2017improved} proposed to add a gradient penalty to keep the Lipschitz continuity, and the loss for the discriminator becomes
\begin{align*}
	\min_{\theta_D} \E_{\mX,\mZ,\tilde{\mX}}-D(\mX;\theta_D)+D(G(\mZ;\theta_G);\theta_D) +\lambda (\Vert \nabla_{\tilde{\mX}}D(\tilde{\mX};\theta_D) \Vert-1)^2
\end{align*}
where $\lambda$ is a constant, $\tilde{\mX}=(1-U)\mX+U\, G(\mZ;\theta_G)$ is a linear interpolation between $\mX$ and $G(\mZ;\theta_G)$, and $U$ follows the uniform distribution over $(0,1)$.
\end{document}